\documentclass[journal]{IEEEtran}

\usepackage[T1]{fontenc}
\usepackage[utf8]{inputenc}
\usepackage{cite}
\usepackage{amsmath,amssymb,amsfonts,bm}
\usepackage{amsthm}
\usepackage{graphicx}
\usepackage{booktabs}
\usepackage{array}
\usepackage{url}
\usepackage{tikz}
\usetikzlibrary{arrows.meta,positioning,fit,calc}
\usepackage[caption=false,font=footnotesize]{subfig}
\usepackage{stfloats}

\usepackage[switch]{lineno}
\usepackage{algpseudocode}
\usepackage{algorithm}
\algnewcommand{\LineComment}[1]{\State \(\triangleright\) #1}
\algtext*{EndFor}
\algtext*{EndIf}
\algtext*{EndWhile}
\algtext*{EndFunction}

\newcommand{\NRMSE}{\operatorname{NRMSE}}

\begin{document}
\bstctlcite{BSTcontrol}

\title{Quaternion-Valued Adaptive Projection Statistics for Neural Receivers in Dual-Polarization Free-Space Quantum Communications}

\author{Anders~M.~Buvarp,~\IEEEmembership{Senior Member,~IEEE,}
        Stefan~Werner,~\IEEEmembership{Fellow,~IEEE}
\thanks{This work was supported by the Research Council of Norway.}
\thanks{A. M. Buvarp and S. Werner are with the Department of Electronic Systems, Norwegian University of Science and Technology, Trondheim, Norway. E-mail: anders.buvarp@ntnu.no. Stefan Werner is also with the Department of Information and Communications Engineering, Aalto University, 00076, Finland.}}

\maketitle

\begin{abstract}
Dual-polarization free-space quantum communication receivers produce two
coupled complex heterodyne streams. We develop robust quaternion receiver
methods for such links under fading, polarization mixing, calibration
mismatch, impulsive contamination, and other non-Gaussian disturbances.
The four polarization and quadrature components are encoded as
quaternion-valued samples. We formulate an identifiable effective channel
model, Quaternion-Valued Projection Statistics (QVPS), and Adaptive QVPS
(AQVPS), which augments samplewise projection outlyingness with robust
local second-order descriptors. Robust complex and quaternion channel
estimators use radial Huber weighting, while QVPS and AQVPS are evaluated
separately as disturbance diagnostics. Neural experiments compare matched
real-, complex-, quaternion-, and widely linear complex-valued receivers.
Robust estimation reduces sensitivity to non-Gaussian contamination.
Local second-order descriptors improve detection of the tested localized
disturbances relative to QVPS, and a descriptor-only ablation slightly
outperforms full AQVPS. Complex LMMSE gives the lowest evaluation MSE in
the clean Gaussian \textrm{SNR} sweep, whereas the Quaternion NN, among
the compared receivers with their nominal activations, gives the lowest
mean evaluation MSE in the tested contaminated sweeps. Activation-matched
experiments indicate that much of the nominal Quaternion NN robustness is
attributable to bounded activation rather than quaternion parameterization
alone. The methods are evaluated at the receiver processing level and do
not constitute a composable CV-QKD security proof.
\end{abstract}


\section{Introduction}
Free-space optical (FSO) and satellite quantum communication can extend quantum-secure links beyond fixed fiber infrastructure, but practical receiver processing remains difficult under realistic propagation and calibration conditions. Continuous-variable quantum key distribution (CV-QKD) is a particularly relevant setting because coherent detection and heterodyne measurements connect naturally to the signal processing and optical communication literature, including dual-polarization receiver architectures and polarization-drift compensation methods \cite{Pereira2023,Guilherme2025,Almeida2025a}. However, free-space and satellite links introduce fading, pointing fluctuations, time-varying transmittance, polarization mixing, residual calibration mismatch, excess noise, and non-Gaussian disturbances that are not well captured by stationary Gaussian receiver models \cite{Muga2024,Clavier2021,Middleton1999}.

In dual-polarization FSO receivers, atmospheric turbulence, pointing
jitter, and geometric loss can produce correlated fading across the
received optical fields~\cite{Jaiswal2018,Kim2026}. These effects occur
together with polarization mixing and receiver mismatch, so the two
heterodyne branches cannot in general be treated as statistically
independent scalar observations. This motivates receiver processing that
acts jointly on the polarization and quadrature components rather than
separately on each branch.

Dual-polarization heterodyne reception produces two complex samples, one
per polarization branch. Equivalently, each receiver sample contains four
real components, $(I_X,Q_X,I_Y,Q_Y)$. These components are not four
unrelated scalar channels. They arise from two coupled complex optical
fields, two polarizations, and two quadratures mixed by a physically
structured receiver channel. Real-valued and augmented complex
formulations can represent the same receiver information. In this work,
we use the quaternion representation to organize the four polarization
and quadrature components jointly and to impose shared observation-level
processing across the coupled components.

Quaternion-valued signal processing provides a compact representation for
four coupled real components. In the present receiver, these components
are $(I_X,Q_X,I_Y,Q_Y)$ and arise from two complex polarization branches.
Building on prior work on quaternion filtering and quaternion neural
networks \cite{Werner2020,Singh2024}, and on our previous work on
polarization-space communications \cite{AndersQPSM,AndersR4C}, we examine
quaternion processing for joint polarization--quadrature receiver
processing. The claim is not that quaternions add dimensionality beyond
$\mathbb{R}^4$. Rather, they provide a structured representation in which
the coupled four-component observation can be processed jointly. This
representation is used here for right-Hamilton neural layers and for
samplewise projection outlyingness over the complete receiver
observation.

The paper focuses on receiver-side robustness. In a CV-QKD application, receiver errors can bias residual noise and other receiver-side quantities used in parameter estimation. In a broader free-space quantum communication setting, the same receiver problems affect channel tracking, polarization demixing, and residual noise stability. We therefore evaluate learned and robust receiver components as signal processing methods. We do not claim that a neural receiver constitutes a security proof.

The main contributions are as follows:\\
\emph{1) Effective channel estimation:} We formulate identifiable
effective channel estimation for the dual-polarization receiver
and compare conventional complex least squares with rowwise
complex and joint quaternion robust weighting under fading and
non-Gaussian contamination.\\
\emph{2) Quaternion encoding and neural receiver:} We encode the two complex
heterodyne outputs as quaternion-valued receiver samples and develop a
right-Hamilton quaternion neural receiver for joint dual-polarization
modulation sample estimation and demixing.\\
\emph{3) QVPS and AQVPS:} We formulate Quaternion-Valued Projection
Statistics (QVPS) for joint quaternion receiver residuals as a
samplewise quaternion outlyingness score. QVPS forms robust scalar directional
projections of the complete four-component quaternion observation and uses
median/MAD standardization to identify samples that are anomalous along at least one joint direction across the polarization and quadrature components. We further develop
Adaptive QVPS (AQVPS), which augments the samplewise QVPS score with robust
local second-order descriptors for I/Q impropriety, cross-polarization
pseudo-covariance, and cross-polarization covariance
\cite{Mili1996,Donoho1992,Resende2024,Huber1964,Hampel1986,Maronna2019,Zhao2024}.\\
\emph{4) Robust estimation and diagnostics:} We develop robust complex
and quaternion effective channel estimation using radial Huber weighting
and compare rowwise complex weighting with shared quaternion weighting.
QVPS and AQVPS are evaluated separately as disturbance diagnostics.\\

Relative to our earlier work on polarization-space modulation with
quaternion neural receivers~\cite{AndersQPSM} and robust communication
with complex and quaternion neural networks~\cite{AndersR4C}, QUARTZ
focuses on robust dual-polarization receiver processing under
non-Gaussian and structured disturbances. The new elements are the
AQVPS fusion of samplewise quaternion outlyingness with local
impropriety, covariance, and pseudo-covariance descriptors, the
comparison of shared quaternion and rowwise complex robust weighting
for effective channel estimation, and matched neural controls that
separate quaternion structure from the effect of bounded activation.

The remainder of this paper is organized as follows.
Sections~\ref{sec:bg}--\ref{sec:training} review the relevant background
and develop the receiver model, quaternion neural receiver, QVPS/AQVPS,
and robust estimation methods. Sections~\ref{sec:simulations} and
\ref{sec:results} describe the simulation setup and numerical results.
Sections~\ref{sec:complexity} and \ref{sec:security} discuss complexity,
implementation, and security limitations, and
Section~\ref{sec:conclusion} concludes the paper.

\noindent \emph{Notation:}
Bold lowercase and uppercase letters denote vectors and matrices,
respectively. The sets of real, complex, and quaternion numbers are
denoted by $\mathbb{R}$, $\mathbb{C}$, and $\mathbb{H}$. The quaternion
units are denoted by $\mathrm{i}$, $\mathrm{j}$, and $\mathrm{k}$, with
$\mathrm{i}^2=\mathrm{j}^2=\mathrm{k}^2=
\mathrm{i}\mathrm{j}\mathrm{k}=-1$.
The Hermitian transpose, complex conjugate, and quaternion conjugate are
denoted by $(\cdot)^H$, $\overline{(\cdot)}$, and $(\cdot)^\ast$,
respectively. The Euclidean and Frobenius norms are denoted by
$\lVert\cdot\rVert_2$ and $\lVert\cdot\rVert_F$, and
$\operatorname{tr}(\cdot)$ and $\mathbb{E}[\cdot]$ denote trace and
expectation. The $n\times n$ identity matrix is $\mathbf{I}_n$.
Estimated quantities are marked with hats.
$\mathcal{O}(\cdot)$ denotes Bachmann--Landau big-$O$ notation, and
$\mathcal{N}_{\mathrm C}(\bm\mu,\bm\Sigma)$ denotes a proper complex
Gaussian distribution.

\section{Related Work}
\label{sec:bg}

CV-QKD has a mature theoretical foundation based on coherent states, Gaussian modulation, and security analyses. This paper focuses on the receiver processing layer rather than on a composable security proof. Free-space and satellite channels complicate the receiver problem because transmittance and noise may vary in time and may be non-Gaussian after fading. Dual-polarization heterodyne architectures can reduce the need for active polarization matching, but receiver DSP is still required to handle polarization drift, imbalance, and residual mixing \cite{Pereira2023,Almeida2025,Guilherme2025,Almeida2025a}.

Classical coherent optical communications routinely use $2\times2$ MIMO
equalization for dual-polarization signals, with the constant modulus
algorithm and related blind equalizers providing practical polarization-tracking baselines~\cite{Godard1980,Savory2008}. More
recent learned equalizers, including VAE-based methods, suggest that
blind or semi-blind learning may help in rapidly varying optical links
\cite{Lauinger2022,Caciularu2020,Qin2026}. For quantum communication,
however, learned equalizers must be evaluated carefully because they may
bias residual noise or residual covariance estimates.

Quaternion-valued signal processing is natural when four coupled real components represent a structured physical sample. In the present problem, the four components correspond to two complex polarizations. Quaternion neural networks have previously been useful for communication signals with coupled components, including polarization-related and electromagnetic information-theory settings \cite{Singh2024,AndersQPSM,AndersR4C}. The contribution here is to connect quaternion neural receivers to a robust free-space quantum receiver model.

Projection statistics and projection outlyingness provide a way to detect
samples that are unusual along some direction of the data space
\cite{Mili1996,Donoho1992,Resende2024,Zhao2024}. Existing real-valued and
complex-valued robust statistics motivate extending this idea to the
coupled four-component receiver representation
\cite{Huber1964,Hampel1986,Maronna2019,Adali2014}. In the proposed QVPS
construction, quaternion-valued receiver samples are projected onto real
scalar directions through $\operatorname{Re}(q u^\ast)$, followed by
robust median/MAD standardization and maximization over a finite set of
unit quaternion directions. This yields a single outlyingness score for
the joint polarization and quadrature observation rather than separate
scores for individual components.

\begin{figure*}[t]
\centering

\noindent\resizebox{\textwidth}{!}{%
\begin{tikzpicture}[
>=latex,
node distance=1.35cm,
block/.style={
draw,
rounded corners,
minimum width=2.7cm,
minimum height=1.4cm,
align=center,
fill=gray!10,
font=\small
},
arrow/.style={
->,
thick
}
]


\node[block] (src)
{
Source\\[2mm]
$\mathbf{x}$
};

\node[block,right=of src] (tx)
{
Dual-Polarization\\
Transmitter
};

\node[block,right=of tx,minimum width=4.2cm] (channel)
{
Free-Space Quantum Channel\\[1mm]
Atmospheric turbulence\\
Polarization mixing\\
Pointing errors\\
Receiver impairments
};

\node[block,right=of channel] (frontend)
{
Dual-Polarization\\
Receiver
};

\node[block,right=of frontend] (quat)
{
Quaternion\\
Encoder
};

\node[block,right=of quat,minimum width=3.0cm] (nn)
{
Quaternion\\
Neural Receiver
};

\node[block,right=of nn] (est)
{
Estimated\\
Modulation\\
Samples
};


\node[block,below=1.45cm of nn] (qvps)
{
QVPS / AQVPS\\
Disturbance Diagnostics
};


\draw[arrow]
(src) --
node[above] {$\mathbf{x}$}
(tx);

\draw[arrow]
(tx) --
node[above] {$\mathbf{s}$}
(channel);

\draw[arrow]
(channel) --
node[above] {$\mathbf{y}$}
(frontend);

\draw[arrow]
(frontend) --
node[above] {$\mathbf{y}$}
(quat);

\draw[arrow]
(quat) --
node[above] {$q_y$}
(nn);

\draw[arrow]
(nn) --
node[above] {$\hat{\mathbf{x}}$}
(est);


\draw[arrow]
(quat) |- (qvps);

\end{tikzpicture}%
}

\caption{Overall processing chain of the robust quaternion receiver.
Dual-polarization observations are encoded as quaternion-valued samples.
The quaternion neural receiver performs joint dual-polarization
modulation sample estimation. QVPS and AQVPS are evaluated separately
as disturbance diagnostics.}

\label{fig:system}
\end{figure*}
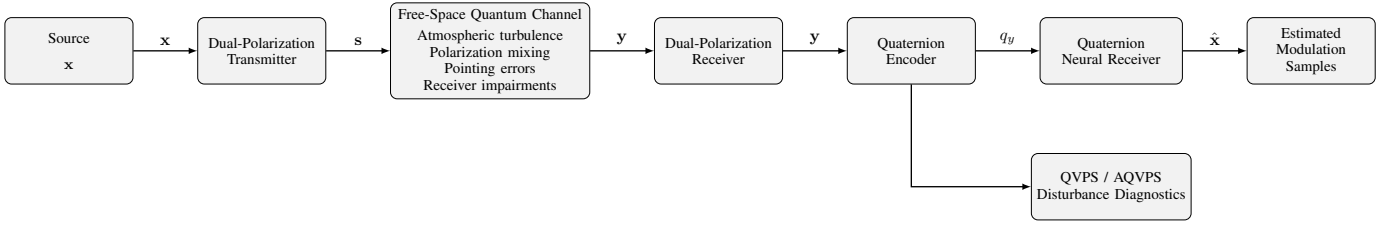

\section{Dual-Polarization Free-Space Quantum System Model}
\label{sec:model}

We consider a dual-polarization free-space quantum communication link with
polarization-diverse coherent reception. The received signal is modeled as
a time-varying $2\times2$ complex MIMO channel that captures polarization
mixing together with free-space propagation and receiver impairments. The
two complex heterodyne outputs are subsequently represented as a single
quaternion-valued observation for joint polarization--quadrature
processing. Fig.~\ref{fig:system} summarizes the resulting receiver
processing chain.

\subsection{Source and Dual-Polarization Transmitter}

The transmitter emits dual-polarization complex modulation samples on the
horizontal and vertical branches. The CV-QKD-oriented simulations use
Gaussian-modulated complex samples unless otherwise stated. A sample is
known to the receiver only when it is assigned as a pilot or subsequently
revealed for receiver-parameter estimation. Unrevealed modulation samples
are not assumed to be receiver labels. Polarization multiplexing and
coherent reception produce the two complex baseband observations used below.

\subsection{Free-Space Quantum Channel}

The transmitted optical fields propagate through a dual-polarization free-space optical channel before coherent reception. During propagation, the optical waveform is impaired by atmospheric turbulence, polarization coupling, beam wandering, pointing errors, phase fluctuations, and receiver noise. Collectively, these impairments introduce amplitude distortion, phase rotation, polarization mixing, and non-Gaussian disturbances that complicate receiver estimation and quadrature recovery.

Atmospheric turbulence produces random refractive-index fluctuations that cause scintillation, wavefront distortion, beam spreading, and angle-of-arrival fluctuations. Depending on the propagation conditions, the turbulence-induced fading may be modeled using log-normal, Gamma--Gamma, or negative exponential distributions. In addition, residual phase fluctuations caused by turbulence and laser linewidth contribute to random carrier phase errors~\cite{Khalighi2014}.

The dual-polarization optical field experiences polarization-dependent distortions due to imperfect optical alignment, propagation through birefringent media, and receiver imperfections. These effects are conveniently represented by a complex $2\times2$ polarization channel matrix
\begin{equation}
\mathbf{H}_k
=\begin{bmatrix}
h_{XX,k} & h_{XY,k}\\
h_{YX,k} & h_{YY,k}
\end{bmatrix},
\end{equation}
where the diagonal elements describe the desired polarization transfer and the off-diagonal elements represent cross-polarization leakage and polarization mixing.

Additional impairments arise from transmitter and receiver hardware, including pointing errors, finite extinction ratios, polarization misalignment, local-oscillator phase noise, shot noise, thermal noise, analog front-end nonlinearities, and imperfect coherent detection. These impairments may be modeled as additive, multiplicative, or coupled disturbances depending on the underlying physical mechanism.

Collecting the two complex polarization components into the quaternion
representation gives
\begin{equation}
q_{y,t}=y_{X,t}+y_{Y,t}\mathrm{j},
\end{equation}
where \(y_{X,t}\) and \(y_{Y,t}\) denote the received complex heterodyne
samples on the two polarization branches. This representation preserves the
joint polarization--quadrature structure and forms the input to the
quaternion receiver developed below.

\subsection{System Model}

At sample time $t$, the transmitter emits the dual-polarization complex
baseband modulation sample
\begin{equation}
    \bm x_t =
    \begin{bmatrix}
        x_{X,t}\\
        x_{Y,t}
    \end{bmatrix}
    \in\mathbb{C}^{2},
\end{equation}
where $x_{X,t}$ and $x_{Y,t}$ denote the transmitted modulation samples on
the horizontal and vertical polarization branches, respectively. In the
reported CV-QKD-oriented simulations, $\bm x_t$ is Gaussian-modulated
unless otherwise stated. It is available to the receiver only for pilot or
revealed-reference estimation blocks, not as a label for all unrevealed
samples.

During free-space propagation, the optical signal experiences atmospheric
turbulence, pointing errors, polarization mixing, phase fluctuations, and
receiver impairments. We model the received dual-polarization sample as
\begin{equation}
    \bm y_t
    =
    \sqrt{T_t}\,
    \bm H_t
    \bm x_t
    +
    \bm n_t,
    \label{eq:system_model}
\end{equation}
where \(T_t>0\) denotes the time-varying channel transmittance,
\(\bm H_t\in\mathbb{C}^{2\times2}\) denotes the dual-polarization
receiver channel, and \(\bm n_t\) denotes the aggregate receiver
disturbance. The off-diagonal elements of \(\bm H_t\) capture
cross-polarization coupling and polarization mixing, while gain imbalance,
residual phase rotation, and calibration mismatch may be incorporated into
its entries. General I/Q imbalance is instead modeled as a separate widely
linear receiver distortion.

The receiver disturbance is modeled as
\begin{equation}
    \bm n_t
    =
    \bm n_t^{(G)}
    +
    \bm n_t^{(\xi)}
    +
    \bm e_t,
    \label{eq:receiver_noise}
\end{equation}
where \(\bm n_t^{(G)}\) denotes calibrated Gaussian receiver noise,
\(\bm n_t^{(\xi)}\) models excess-noise contributions, and \(\bm e_t\)
represents impulsive, heavy-tailed, contaminated, or other non-Gaussian
disturbances.

We write
\begin{equation}
    T_t = \bar T f_t,
\end{equation}
where \(\bar T\) is the mean transmittance and \(f_t>0\) is a fading
process. In the simulations, \(f_t\) is specified according to the
corresponding propagation scenario.

\subsection{Quaternion Observation Model}

The dual-polarization coherent receiver performs polarization-diverse coherent detection and produces the complex heterodyne observations
\begin{equation}
    y_{X,t}=I_{X,t}+\mathrm{i}Q_{X,t},
\qquad
y_{Y,t}=I_{Y,t}+\mathrm{i}Q_{Y,t},
\end{equation}
which are collected as
\begin{equation}
    \bm y_t
    =
    \begin{bmatrix}
        y_{X,t}\\
        y_{Y,t}
    \end{bmatrix}
    \in\mathbb{C}^{2}.
\end{equation}
Using the Cayley–Dickson construction~\cite{AndersR4C} for quaternion-valued processing, the observation is embedded into
\begin{equation}
q_{y,t}
=
y_{X,t}
+
y_{Y,t}\mathrm{j},
\end{equation}
which preserves the joint polarization structure and forms the input to the proposed robust quaternion neural receiver. The quaternion encoding is
\begin{equation}
    q_{y,t}
=
y_{X,t}+y_{Y,t}\mathrm{j}
=
I_{X,t}
+\mathrm{i}Q_{X,t}
+\mathrm{j}I_{Y,t}
+\mathrm{k}Q_{Y,t},
\label{eq:q_encoding}
\end{equation}
with the same convention for $q_{x,t}$. The units satisfy
$\mathrm{i}^2=\mathrm{j}^2=\mathrm{k}^2
=\mathrm{i}\mathrm{j}\mathrm{k}=-1$. This encoding treats the four heterodyne components as one coupled object. The complex $2\times2$ MIMO model in \eqref{eq:system_model} can be represented using an augmented or bilateral quaternion operator acting on $q_{x,t}$ \cite{CheongTook2011,CheongTook2010}. For the structured Jones matrix considered below, the mapping reduces to a single right-Hamilton multiplication.

\section{Robust Quaternion Neural Receiver}
\label{sec:qnr}

A quaternion neural receiver maps the received quaternion sample to an
estimate of the transmitted dual-polarization modulation sample,
\begin{equation}
    \widehat q_{x,t}
    =
    f_{\bm\Theta}(q_{y,t}),
    \label{eq:qnn_mapping}
\end{equation}
where $f_{\bm\Theta}$ is a quaternion-valued neural network with
trainable parameters $\bm\Theta$. In the numerical experiments, the
network is evaluated for joint dual-polarization modulation sample estimation and
demixing. Extension of the neural mapping to channel or receiver-parameter
estimation is not evaluated here.

For the structured Jones subfamily in \eqref{eq:jones_matrix}, the
quaternion encoding represents the dual-polarization transformation by
right-Hamilton multiplication. The receiver therefore uses right-sided
quaternion linear layers
\begin{equation}
    \bm h_{\ell+1}
    =
    \phi_{\ell}
    \left(
        \bm h_{\ell}\bm W_{\ell}
        +
        \bm b_{\ell}
    \right),
    \label{eq:right_quaternion_layer}
\end{equation}
where $\bm W_{\ell}$ and $\bm b_{\ell}$ are quaternion-valued weights
and biases, and $\phi_{\ell}$ is a quaternion compatible activation.

For the structured Jones matrix
\begin{equation}
    \bm H
    =
    \begin{bmatrix}
    \alpha & -\overline{\beta}\\
    \beta  & \overline{\alpha}
\end{bmatrix},
    \label{eq:jones_matrix}
\end{equation}
the corresponding quaternion mapping under the encoding in \eqref{eq:q_encoding} is
\begin{equation}
    q_y
    =
    q_x w,
    \qquad
    w
=
\Re\{\alpha\}
+
\Im\{\alpha\}\mathrm{i}
+
\Re\{\beta\}\mathrm{j}
+
\Im\{\beta\}\mathrm{k}.
    \label{eq:jones_right_hamilton}
\end{equation}
Thus, the right-Hamilton layer preserves the multiplication convention
induced by the structured Jones representation rather than imposing the
generally different left-sided mapping $\bm W_{\ell}\bm h_{\ell}$. This
identity does not represent an arbitrary complex $2\times2$ MIMO
operator, which requires an augmented or bilateral quaternion
representation. Outside the structured Jones setting, a right-Hamilton
layer is a structured inductive bias.

Given an effective channel estimate, define the calibrated residual
\begin{equation}
    \bm r_t
    =
    \bm y_t-\widehat{\bm A}\bm x_t.
    \label{eq:calibrated_residual}
\end{equation}

\section{Quaternion-Valued Projection Statistics}
\label{sec:qvps}

Projection statistics identify multivariate observations that exhibit a
large standardized deviation along at least one direction in the data
space. For quaternion-valued receiver observations, we retain the coupled
four-component representation while forming real directional projections.
Let
\begin{equation}
q_i
=
q_{i,0}
+
q_{i,1}\mathrm{i}
+
q_{i,2}\mathrm{j}
+
q_{i,3}\mathrm{k}
\in\mathbb{H}
\end{equation}
denote a quaternion-valued receiver sample or residual, with corresponding
real representation
\begin{equation}
\widetilde{\bm q}_i
=
\begin{bmatrix}
q_{i,0} &
q_{i,1} &
q_{i,2} &
q_{i,3}
\end{bmatrix}^{T}
\in\mathbb{R}^{4}.
\end{equation}

For a unit quaternion direction
\begin{equation}
u_\ell
=
u_{\ell,0}
+
u_{\ell,1}\mathrm{i}
+
u_{\ell,2}\mathrm{j}
+
u_{\ell,3}\mathrm{k},
\qquad
|u_\ell|_{\mathbb H}=1,
\end{equation}
define the scalar directional projection
\begin{equation}
z_{i\ell}
=
\operatorname{Re}\!\left(q_i u_\ell^\ast\right)
=
\widetilde{\bm q}_i^{\,T}\widetilde{\bm u}_\ell,
\label{eq:qvps_projection}
\end{equation}
where
\begin{equation}
\widetilde{\bm u}_\ell
=
\begin{bmatrix}
u_{\ell,0} &
u_{\ell,1} &
u_{\ell,2} &
u_{\ell,3}
\end{bmatrix}^{T}.
\end{equation}
Thus, each projection acts on the complete quaternion observation and
tests a joint direction across the two polarization and quadrature
components.

For projection direction $\ell$, the robust location and scale estimates are
\begin{equation}
m_\ell
=
\operatorname{med}_{i}
\left\{
z_{i\ell}
\right\},
\label{eq:qvps_location}
\end{equation}
and
\begin{equation}
\widehat{s}_\ell
=
1.4826\,
\operatorname{med}_{i}
\left\{
\left|z_{i\ell}-m_\ell\right|
\right\},
\label{eq:qvps_scale}
\end{equation}
where $1.4826$ gives the usual normal-consistent MAD scale estimate.
The standardized directional deviation is
\begin{equation}
d_{i\ell}
=
\frac{
\left|z_{i\ell}-m_\ell\right|
}{
\widehat{s}_\ell+\epsilon_s
},
\label{eq:qvps_directional_deviation}
\end{equation}
where $\epsilon_s$ is a small numerical constant used only to avoid
division by zero. In the numerical implementation,
$\epsilon_s=10^{-12}$.

The proposed Quaternion-Valued Projection Statistics (QVPS) score is the
largest standardized deviation over the evaluated projection directions,
\begin{equation}
O_{\mathbb H}(q_i)
=
\max_{\ell=1,\ldots,L}
d_{i\ell}.
\label{eq:qvps}
\end{equation}
The finite-direction maximum approximates the supremum used in classical
projection outlyingness. Although the individual projections in
\eqref{eq:qvps_projection} are real-valued, they are formed from the
complete quaternion observation. QVPS therefore provides a single
samplewise outlyingness score for the coupled
$(I_X,Q_X,I_Y,Q_Y)$ receiver components rather than separate scores for
individual scalar or polarization components. Under the quaternion--
$\mathbb{R}^{4}$ isometry, and for the same finite set of directions,
\eqref{eq:qvps_projection}--\eqref{eq:qvps} are exactly
finite-direction real $\mathbb{R}^{4}$ projection outlyingness. The
quaternion notation does not define a new scalar projection statistic.
It provides a joint polarization--quadrature representation, one
observation-level weight, and compatibility with the structured receiver
representation in Section~\ref{sec:qnr}.

The projection directions are sampled independently and uniformly on the
unit quaternion sphere $S^3$. Specifically,
\begin{equation}
u_\ell
=
\frac{
a_\ell
+
b_\ell\mathrm{i}
+
c_\ell\mathrm{j}
+
d_\ell\mathrm{k}
}{
\sqrt{
a_\ell^2+b_\ell^2+c_\ell^2+d_\ell^2
}
},
\label{eq:qvps_direction}
\end{equation}
where
\begin{equation}
a_\ell,b_\ell,c_\ell,d_\ell
\overset{\mathrm{i.i.d.}}{\sim}
\mathcal N(0,1).
\end{equation}

\begin{algorithm}[t!]
\caption{Quaternion-Valued Projection Statistics (QVPS).}
\label{alg:qvps}
\textbf{Input:} Quaternion observations
$\mathcal{Q}=\{q_1,\ldots,q_N\}$, number of projection directions $L$,
and weighting constant $c_Q>0$\\
\textbf{Output:} Projection outlyingness
$\{O_{\mathbb H}(q_i)\}_{i=1}^{N}$ and robust sample weights
$\{\omega_i\}_{i=1}^{N}$
\begin{algorithmic}[1]

\For{$\ell=1$ \textbf{to} $L$}
    \State Draw a unit quaternion direction $u_\ell$ uniformly on $S^3$

    \For{$i=1$ \textbf{to} $N$}
        \State
        $z_{i\ell}
        \gets
        \operatorname{Re}\!\left(q_i u_\ell^\ast\right)$
    \EndFor

    \State
    $m_\ell
    \gets
    \operatorname{med}_{i}\{z_{i\ell}\}$

    \State
    $s_\ell
    \gets
    1.4826\,
    \operatorname{med}_{i}
    \{|z_{i\ell}-m_\ell|\}$

    \For{$i=1$ \textbf{to} $N$}
        \State
        $d_{i\ell}
        \gets
        |z_{i\ell}-m_\ell|/(s_\ell+\epsilon_s)$
    \EndFor
\EndFor

\For{$i=1$ \textbf{to} $N$}
    \State
    $O_{\mathbb H}(q_i)
    \gets
    \max_{\ell=1,\ldots,L} d_{i\ell}$

    \State
$\omega_i
\gets
\min\left\{
1,\,
\frac{c_Q}
{O_{\mathbb H}(q_i)+\epsilon_w}
\right\}$
\EndFor

\State \Return
$\{O_{\mathbb H}(q_i)\}_{i=1}^{N}$,
$\{\omega_i\}_{i=1}^{N}$

\end{algorithmic}
\end{algorithm}
In practice, $L$ randomly sampled directions are used. The resulting
outlyingness score is converted to a robust sample weight using
\begin{equation}
\omega_i
=
\min\left\{
1,\,
\frac{c_Q}
{O_{\mathbb H}(q_i)+\epsilon_w}
\right\},
\label{eq:qvps_weight}
\end{equation}
where $c_Q>0$ controls the onset of downweighting and $\epsilon_w$ is a
small numerical constant used to avoid division by zero. In the numerical
experiments, $c_Q=2.7$ and $\epsilon_w=10^{-12}$. Samples with
$O_{\mathbb H}(q_i)\lesssim c_Q$ retain unit weight, while samples with
larger outlyingness scores are progressively downweighted. The resulting QVPS weights define an observation-level weighting rule.
The QVPS outlyingness score also constitutes the samplewise descriptor
incorporated into AQVPS in the following subsection. In the reported experiments, QVPS and
AQVPS are evaluated as disturbance diagnostics, while the robust channel
estimation results use radial Huber weights.

The complete QVPS procedure is summarized in
Algorithm~\ref{alg:qvps}. In practice, the full-direction supremum is approximated by evaluating a finite set of $L$
randomly sampled unit quaternion directions.

\begin{algorithm}[b!]
\caption{Adaptive Quaternion Projection Statistics (AQVPS) via Robust
Multivariate Descriptor Fusion.}
\label{alg:aqvps}
\textbf{Input:} Quaternion residuals
$\mathcal Q=\{q_1,\ldots,q_N\}$, QVPS outlyingness
$\{O_{\mathbb H}(q_i)\}_{i=1}^{N}$, window length $W$,
and clean-calibrated weighting constant $c_A>0$\\
\textbf{Output:} AQVPS outlyingness
$\{O_{\mathrm A}(q_i)\}_{i=1}^{N}$ and adaptive sample weights
$\{\omega_i\}_{i=1}^{N}$
\begin{algorithmic}[1]

\For{$i=1$ \textbf{to} $N$}
    \State Form the symmetric, edge-shifted local window $\mathcal W_i$
    \State Estimate winsorized local powers $P_X,P_Y$ over $\mathcal W_i$
    \State Estimate robust complex $C_{XY},P_{XY},K_X,K_Y$ over $\mathcal W_i$
    \State Form and clip $\rho_C,\rho_P,\kappa_X,\kappa_Y$
    \State Construct $\mathbf d_i$ from $O_{\mathbb H}(q_i)$, the
    local power ratio, $\rho_C$, $\rho_P$, $\kappa_X$, and $\kappa_Y$
    \State Apply the log and clipped-atanh descriptor transforms
\EndFor

\State Estimate robust descriptor location
$\widehat{\boldsymbol\mu}_{\mathrm A}$
\State Estimate robust descriptor scatter
$\widehat{\mathbf\Sigma}_{\mathrm A}$

\For{$i=1$ \textbf{to} $N$}
    \State $O_{\mathrm A}(q_i)
    \gets
    \sqrt{
    (\mathbf d_i-\widehat{\boldsymbol\mu}_{\mathrm A})^{T}
    \widehat{\mathbf\Sigma}_{\mathrm A}^{-1}
    (\mathbf d_i-\widehat{\boldsymbol\mu}_{\mathrm A})
    }$
    \State
$\omega_i
\gets
\min\left\{
1,\,
\frac{c_A}
{O_{\mathrm A}(q_i)+\epsilon_w}
\right\}$
\EndFor

\State \Return
$\{O_{\mathrm A}(q_i)\}_{i=1}^{N}$,
$\{\omega_i\}_{i=1}^{N}$

\end{algorithmic}
\end{algorithm}

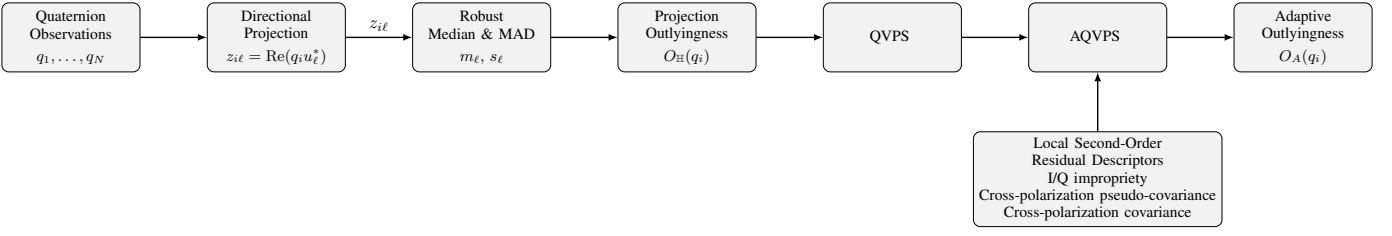
\begin{figure*}[t!]
\centering

\noindent\resizebox{\textwidth}{!}{%
\begin{tikzpicture}[
>=latex,
node distance=1.35cm,
block/.style={
draw,
rounded corners,
minimum width=2.8cm,
minimum height=1.4cm,
align=center,
fill=gray!10,
font=\small
},
arrow/.style={->,thick}
]


\node[block] (obs)
{
Quaternion\\
Observations\\[1mm]
$q_1,\ldots,q_N$
};

\node[block,right=of obs] (proj)
{
Directional\\
Projection\\[1mm]
$z_{i\ell}=\operatorname{Re}(q_i u_\ell^\ast)$
};

\node[block,right=of proj] (scale)
{
Robust\\
Median \& MAD\\[1mm]
$m_\ell,\,s_\ell$
};

\node[block,right=of scale] (out)
{
Projection\\
Outlyingness\\[1mm]
$O_{\mathbb H}(q_i)$
};

\node[block,right=of out] (qvps)
{
QVPS
};

\node[block,right=of qvps] (aqvps)
{
AQVPS
};

\node[block,right=of aqvps] (weights)
{
Adaptive\\
Outlyingness\\[1mm]
$O_A(q_i)$
};


\draw[arrow] (obs) -- (proj);

\draw[arrow] (proj) -- node[above]
{$z_{i\ell}$}
(scale);

\draw[arrow] (scale) -- (out);

\draw[arrow] (out) -- (qvps);

\draw[arrow] (qvps) -- (aqvps);

\draw[arrow] (aqvps) -- (weights);


\node[block,below=1.2cm of aqvps,
minimum width=4.4cm]
(extra)
{
Local Second-Order\\
Residual Descriptors\\
I/Q impropriety\\
Cross-polarization pseudo-covariance\\
Cross-polarization covariance
};

\draw[arrow]
(extra.north) -- (aqvps.south);

\end{tikzpicture}%
}

\caption{Detailed processing chain of QVPS and AQVPS. Quaternion
observations are mapped to real-valued directional projections using
$z_{i\ell}=\operatorname{Re}(q_i u_\ell^\ast)$. Robust median and MAD
estimates are used to compute the QVPS outlyingness score. AQVPS combines
this samplewise score with local second-order residual descriptors to
detect coupled polarization impairments. The resulting scores are
evaluated as disturbance diagnostics.}

\label{fig:qvps}
\end{figure*}

\subsection{Adaptive Quaternion Projection Statistics via Robust Descriptor Fusion}

Adaptive Quaternion Projection Statistics (AQVPS) augments samplewise
QVPS with local second-order residual descriptors. For sample $t$, the
implementation forms a nominally symmetric window $\mathcal W_t$ of
length $W$. At a boundary, the window is shifted inward to retain $W$
samples whenever possible. The local powers $P_X$ and $P_Y$ are
upper-tail winsorized means of $|r_X|^2$ and $|r_Y|^2$. The cap is the sample median plus $2.5$ times a MAD-based robust scale,
using the factor $1.4826$ as a fixed scaling convention. The
cross-polarization covariance $C_{XY}$, pseudo-covariance $P_{XY}$, and
pseudo-moments $K_X,K_Y$ are robust complex radial Huber location
estimates of $r_X\overline{r_Y}$, $r_Xr_Y$, $r_X^2$, and $r_Y^2$,
respectively. These estimates use tuning constant $1.5$, a
coordinatewise-median initialization, and fixed radial scale
$\operatorname{med}|s-\theta|/1.1774+\varepsilon$. They use at most 50
updates and stop when the location change is below
$10^{-6}(\sigma+|\theta|+\varepsilon)$. Their magnitudes enter
\[
\begin{aligned}
\rho_C&=
\frac{|C_{XY}|}
{\sqrt{P_XP_Y+\varepsilon}},
&
\rho_P&=
\frac{|P_{XY}|}
{\sqrt{P_XP_Y+\varepsilon}},
\\
\kappa_X&=
\frac{|K_X|}
{P_X+\varepsilon},
&
\kappa_Y&=
\frac{|K_Y|}
{P_Y+\varepsilon}.
\end{aligned}
\]
The AQVPS descriptor is
\begin{equation}
\mathbf d_t =
\begin{bmatrix}
\log(1+O_{\mathbb H,t})\\
\log\!\left((P_X+\varepsilon)/(P_Y+\varepsilon)\right)\\
\operatorname{atanh}(\bar\rho_C)\\
\operatorname{atanh}(\bar\rho_P)\\
\operatorname{atanh}(\bar\kappa_X)\\
\operatorname{atanh}(\bar\kappa_Y)
\end{bmatrix}.
\label{eq:aqvps_descriptor}
\end{equation}
Here, each barred quantity is clipped to $[0,1-10^{-6}]$ before the
inverse hyperbolic tangent, and $\varepsilon=10^{-12}$ handles
zero-power and division-by-zero cases. Descriptor location is the
coordinatewise median. The reported structured experiment uses diagonal
MAD scatter with scale floor $10^{-6}$, ridge regularization $10^{-3}$,
and maximum condition number $10^{6}$. The clean-data calibration sets
\(c_A\) to the 97.5th percentile of descriptor distances pooled from
three clean calibration blocks and freezes only this threshold for the
disturbance sweep. Descriptor location and regularized diagonal scatter
are re-estimated from the current block at each IRLS iteration.

In the structured-detection experiments, QVPS, AQVPS, and
descriptor-only AQVPS each update their own effective-channel estimate
using weighted least squares with weights derived from their current
method-specific residuals. The detector score reported for each method
is the score from the last executed outer iteration, computed from the
residual immediately preceding the final channel update. The
augmented-covariance baseline instead uses a single ordinary-LS channel
estimate followed by one residual-scoring pass.

For the reported structured experiment,
three clean calibration blocks gave $c_A=4.033$. AQVPS then
transforms the descriptor distance into an adaptive sample weight. Thus,
AQVPS detects localized changes in second-order structure that need not
appear as samplewise amplitude outliers. 
The complete AQVPS procedure is summarized in
Algorithm~\ref{alg:aqvps}. Fig.~\ref{fig:qvps} summarizes the QVPS/AQVPS processing chain,
including the local second-order descriptors used by AQVPS.

\section{Training and Estimation Procedure}
\label{sec:training}

The receiver combines supervised neural estimation with robust
parameter estimation. QVPS and AQVPS are evaluated separately as
disturbance diagnostics. This section specifies the neural training objectives, QVPS and AQVPS
diagnostics, and the corresponding robust receiver estimation procedure.

\subsection{Neural Receiver Training}

Let $\{(\bm y_t,\bm x_t)\}_{t=1}^{N}$ denote received
dual-polarization observations and their known pilot or revealed
modulation samples. Each observation $\bm y_t\in\mathbb{C}^{2}$
is encoded as the quaternion sample
\begin{equation}
    q_{y,t}
    =
    y_{X,t}+y_{Y,t}\mathbf{j},
\end{equation}
with the corresponding reference represented by
$q_{x,t}=x_{X,t}+x_{Y,t}\mathrm{j}$. The quaternion neural receiver
$f_{\bm\Theta}$ is trained to estimate the desired receiver quantity
from the coupled polarization and quadrature components.

For recovery of a dual-polarization modulation sample, the receiver output is
\begin{equation}
    \widehat q_{x,t}
    =
    f_{\bm\Theta}(q_{y,t}),
\end{equation}
and the corresponding quaternion residual is
\begin{equation}
    q_{r,t}
    =
    q_{x,t}-\widehat q_{x,t}.
    \label{eq:training_residual}
\end{equation}
Since the quaternion norm is equal to the Euclidean norm of its four
real components, the nominal training objective can be written as
\begin{equation}
    \mathcal{L}_{0}(\bm\Theta)
    =
    \frac{1}{4N}
    \sum_{t=1}^{N}
    \left\|
        q_{x,t}-f_{\bm\Theta}(q_{y,t})
    \right\|_{\mathbb H}^{2}.
    \label{eq:nominal_training_loss}
\end{equation}
This objective provides the non-robust reference training rule and is
also used for clean-data receiver training in experiments that evaluate
robustness to previously unseen receiver contamination.

For the widely linear Complex NN, the receiver uses the augmented complex
observation
\[
    \underline{\bm y}_t
=
\begin{bmatrix}
    \bm y_t\\
    \overline{\bm y}_t
\end{bmatrix}.
\]
The reported parameter count includes the trainable scalar offsets in
the two modReLU activations.

All neural receivers use three biased affine layers and a linear output.
The architectures and corresponding real-scalar parameter counts are
listed in Table~\ref{tab:neural_architectures}. The parameter counts
differ by at most 1.6\%.

For the fixed-channel neural experiments, we use
\begin{equation}
    \bm H(\theta,\phi,\kappa)
    =
    \begin{bmatrix}
        \cos\theta
        &
        \kappa\sin\theta\,e^{\mathrm{i}\phi}
        \\
        -\kappa\sin\theta\,e^{-\mathrm{i}\phi}
        &
        \cos\theta
    \end{bmatrix},
    \label{eq:neural_jones_generator}
\end{equation}
with
\(
(\theta,\phi,\kappa,\eta)
=
(0.55,0.70,0.45,0.35)
\).
The received signal is generated as
\begin{equation}
    \bm y_t
    =
    \sqrt{\eta}\,
    \bm H(\theta,\phi,\kappa)\bm x_t
    +
    \bm n_t .
\end{equation}

A separate neural receiver is trained for each SNR, contamination
condition, and paired trial. For the contaminated SNR sweeps, training
uses Gaussian-clean observations at the corresponding SNR, while
impulsive contamination is introduced only during evaluation.
\begin{table}[b!]
\caption{Neural Receiver Architectures}
\label{tab:neural_architectures}
\centering
\scriptsize
\begin{tabular}{lccc}
\toprule
Receiver & Dimensions & Activation & Parameters \\
\midrule
Real NN
& $4$--$47$--$47$--$4$
& GELU
& 2,683 \\
Complex NN
& $2$--$34$--$34$--$2$
& modReLU
& 2,726 \\
Quaternion NN
& $1$--$24$--$24$--$1$
& split $\tanh$
& 2,692 \\
widely linear Complex NN
& $4$--$33$--$33$--$2$
& modReLU
& 2,712 \\
\bottomrule
\end{tabular}
\end{table}

QVPS or AQVPS weights could also be incorporated into neural training by
weighting each quaternion residual jointly across its four real components.
We do not evaluate such weighted neural training here.

\subsection{Robust Receiver Parameter Estimation}

We consider model-based estimation of a locally constant effective
channel. Let $\bm x_t\in\mathbb{C}^{2}$ denote a known transmitted or
pilot sample and let $\bm y_t\in\mathbb{C}^{2}$ denote the corresponding
received dual-polarization observation. Define
\begin{equation}
    \bm A=\sqrt{T}\,\bm H.
    \label{eq:effective_channel}
\end{equation}
The locally constant receiver model is
\begin{equation}
    \bm y_t
    =
    \bm A\bm x_t+\bm n_t.
    \label{eq:training_channel_model}
\end{equation}
Without an explicit normalization of $\bm H$, the factors $T$ and
$\bm H$ are not jointly identifiable from these observations. The
estimators below therefore target $\bm A$.
Complex least squares provides the non-robust reference estimate,
\begin{equation}
    \widehat{\bm A}_{\mathrm{LS}}
    =
    \arg\min_{\bm A}
    \sum_{t=1}^{N}
    \left\|
        \bm y_t-\bm A\bm x_t
    \right\|_2^2.
    \label{eq:complex_ls_estimation}
\end{equation}
Under non-Gaussian contamination, a small number of large residuals can
dominate this objective. The robust complex estimator uses separate
rowwise weights for the two complex receiver outputs,
\begin{align}
    \widehat{\bm A}_{\mathrm C}
    &=
    \arg\min_{\bm A}
    \sum_{t=1}^{N}
    \Big[
        \omega_{X,t}
        \left|[\bm y_t-\bm A\bm x_t]_X\right|^2
        \nonumber\\
    &\qquad\qquad\qquad\qquad
        +
        \omega_{Y,t}
        \left|[\bm y_t-\bm A\bm x_t]_Y\right|^2
    \Big],
    \label{eq:weighted_complex_estimation}
\end{align}
where $\omega_{X,t}$ and $\omega_{Y,t}$ are obtained from the
corresponding complex residual magnitudes. The quaternion estimator
instead uses one shared observation weight derived from the joint
residual,
\begin{equation}
    \widehat{\bm A}_{\mathbb H}
    =
    \arg\min_{\bm A}
    \sum_{t=1}^{N}
    \omega_t
    \left\|
        \bm y_t-\bm A\bm x_t
    \right\|_2^2,
    \label{eq:weighted_quaternion_estimation}
\end{equation}
where $\omega_t$ is obtained from the joint four-real-dimensional
residual norm.

For a current effective channel estimate, define
\begin{equation}
    \bm r_t
    =
    \bm y_t-\widehat{\bm A}\bm x_t .
    \label{eq:parameter_estimation_residual}
\end{equation}
Let
\[
    \bm X
    =
    \left[
        \bm x_1,\ldots,\bm x_N
    \right],
    \qquad
    \bm Y
    =
    \left[
        \bm y_1,\ldots,\bm y_N
    \right],
\]
and let \(\bm y_p\) denote row \(p\) of \(\bm Y\), for
\(p\in\{X,Y\}\). For robust complex MIMO estimation, each output row is treated separately.
For \(p\in\{X,Y\}\), define
\begin{align}
    u_{p,t}^{(k)}
    &=
    \left|r_{p,t}^{(k)}\right|,
    \\
    \widehat{\sigma}_{p}^{(k)}
    &=
    \frac{
        \operatorname{median}_{t} u_{p,t}^{(k)}
    }{
        m_{2}^{\mathrm{WH}}+10^{-12}
    }
    +10^{-12},
\end{align}
where \(m_{2}^{\mathrm{WH}}=1.185185\) is the Wilson--Hilferty
approximation to the median of the \(\chi_2\) distribution. The corresponding
Huber weight is
\begin{equation}
    \omega_{p,t}^{(k)}
    =
    \min\left\{
        1,
        \frac{
            c_{\mathrm C}\widehat{\sigma}_{p}^{(k)}
        }{
            u_{p,t}^{(k)}+10^{-12}
        }
    \right\},
    \qquad
    c_{\mathrm C}=2.734139791.
\end{equation}
With
\[
    \bm W_{p}^{(k)}
    =
    \operatorname{diag}
    \left(
        \omega_{p,1}^{(k)},\ldots,\omega_{p,N}^{(k)}
    \right),
\]
the \(p\)th output row, \(p\in\{X,Y\}\), is updated as
\[
    \bm a_{p}^{(k+1)}
    =
    \bm y_{p}
    \bm W_{p}^{(k)}
    \bm X^{H}
    \left(
        \bm X
        \bm W_{p}^{(k)}
        \bm X^{H}
        +
        10^{-8}\bm I_{2}
    \right)^{-1}.
\]

Quaternion robust estimation instead uses the joint residual magnitude
\begin{equation}
    u_{t}^{(k)}
    =
    \sqrt{
        \left|r_{X,t}^{(k)}\right|^{2}
        +
        \left|r_{Y,t}^{(k)}\right|^{2}
    }.
\end{equation}
The corresponding shared scale is
\begin{equation}
    \widehat{\sigma}_{\mathbb H}^{(k)}
    =
    \frac{
        \operatorname{median}_{t} u_{t}^{(k)}
    }{
        m_{4}^{\mathrm{WH}}+10^{-12}
    }
    +10^{-12},
\end{equation}
where \(m_{4}^{\mathrm{WH}}=1.835670\). The shared observation weight is
\begin{equation}
    \omega_{t}^{(k)}
    =
    \min\left\{
        1,
        \frac{
            c_{\mathbb H}\widehat{\sigma}_{\mathbb H}^{(k)}
        }{
            u_{t}^{(k)}+10^{-12}
        }
    \right\},
    \qquad
    c_{\mathbb H}=3.344609346.
\end{equation}
With
\[
    \bm W^{(k)}
    =
    \operatorname{diag}
    \left(
        \omega_{1}^{(k)},\ldots,\omega_{N}^{(k)}
    \right),
\]
the effective channel update is
\[
    \widehat{\bm A}^{(k+1)}
    =
    \bm Y
    \bm W^{(k)}
    \bm X^{H}
    \left(
        \bm X
        \bm W^{(k)}
        \bm X^{H}
        +
        10^{-8}\bm I_{2}
    \right)^{-1}.
\]

The constants \(c_{\mathrm C}=2.734139791\) and
\(c_{\mathbb H}=3.344609346\) are chosen to give the same nominal
\(97.5\%\) clean Gaussian onset under the implemented Wilson--Hilferty
scale normalization.

The robust complex estimator assigns separate radial Huber weights to
the two complex output rows, whereas the quaternion estimator assigns
one shared weight to the joint four-real-dimensional residual. Both
estimators initialize with ridge-stabilized complex LS and stop when the
relative effective-matrix change is below \(10^{-7}\) or the iteration
limit is reached.

For quaternion processing,
\[
q_{r,t}=r_{X,t}+r_{Y,t}\mathrm{j},
\qquad
|q_{r,t}|_{\mathbb H}^{2}
=
|r_{X,t}|^{2}+|r_{Y,t}|^{2}
=
\|\bm r_t\|_{2}^{2}.
\]
Thus, quaternion radial Huber weighting is equivalent to joint
complex-vector radial weighting of the two receiver outputs. Its
distinction from the robust complex estimator is the use of one shared
observation weight rather than separate rowwise weights. QVPS and AQVPS
provide alternative joint residual scores evaluated separately as
disturbance diagnostics.

\subsection{Receiver Operation and Adaptation}

After training or calibration, the receiver uses either the trained
neural mapping or the estimated effective channel and residual
statistics. QVPS and AQVPS are used here as receiver diagnostics computed from
current residuals. Their use for weighted receiver adaptation is not
evaluated in the reported experiments. Blind adaptation would require
an additional receiver-consistency or decision-directed criterion.

\begin{table*}[t!]
\caption{Completed Simulation Configurations}
\label{tab:reproducibility}
\centering
\scriptsize
\begin{tabular}{p{0.18\textwidth}p{0.22\textwidth}p{0.52\textwidth}}
\toprule
Experiment & Trials and seeds & Fixed configuration \\
\midrule

Impulsive contamination estimation
& 500 paired trials, seed 7
& $N=3{,}000$, proper complex Gaussian modulation with variance $4$,
independent fading
$T_t=\operatorname{clip}\{\exp(\mu_T+0.35Z_t),10^{-4},1\}$,
$Z_t\sim\mathcal N(0,1)$,
$\mu_T=\log(0.35)-0.35^2/2$, noise variance $0.02$,
impulsive variance $12^2\times0.02$, mixing angle $0.35$,
dimension-matched Huber constants $c_{\mathrm C}=2.734$ and
$c_{\mathbb H}=3.345$, 20 IRLS iterations. \\

Structured QVPS/AQVPS
& 30 paired trials, simulation seed 811, QVPS seed 812
& $N=3{,}000$, $L=64$ directions, $c_Q=2.7$, $W=151$, maximum 10 outer channel estimation reweighting iterations,
strengths $0,0.1,0.2,0.4,0.6$, affected fraction $0.25$, three clean
calibration blocks, 2{,}000 bootstrap resamples, diagonal AQVPS scatter. \\

Matched neural nominal
& 20 paired trials, seeds 1201/2201/3201
& Train/test sizes 4096/2048, noise variance $0.02$, 100 epochs,
batch size 256, AdamW with learning rate $10^{-3}$ and weight decay
$10^{-5}$. \\

Neural impairment sweeps
& 20 paired trials, seeds 4201/5201/6201
& Train/test sizes 4096/2048, 100 epochs, batch size 256, AdamW
with learning rate $10^{-3}$ and weight decay $10^{-5}$, \textrm{SNR} values
$-5,0,5,10,15,20$~dB, \textrm{SNR} contamination probabilities
$0,0.10,0.15,0.20,0.30,0.45$, fixed test impulse variance $2.88$.
The activation ablation uses 10 trials and 10,000 bootstrap resamples. \\

\bottomrule
\end{tabular}
\end{table*}

\section{Simulation Benchmark}
\label{sec:simulations}

The simulation benchmark evaluates the receiver under controlled
dual-polarization channel and receiver impairments generated according
to \eqref{eq:system_model}. The completed simulation configurations are
summarized in Table~\ref{tab:reproducibility}. Unless otherwise stated,
transmitted samples are Gaussian-modulated complex samples, and paired
Monte Carlo trials use common channel and disturbance realizations
across competing receivers.

The principal model-based performance metric is effective channel NRMSE,
\begin{equation}
    \NRMSE_A
    =
    \frac{
        \|\widehat{\bm A}-\bm A_{\star}\|_F
    }{
        \|\bm A_{\star}\|_F
    }.
\end{equation}
For the fading simulations,
\begin{equation}
    \bm A_{\star}
    =
    \left(
        \sum_{t=1}^{N}
        \sqrt{T_t}\,
        \bm H
        \bm x_t\bm x_t^{H}
    \right)
    \left(
        \sum_{t=1}^{N}
        \bm x_t\bm x_t^{H}
    \right)^{-1},
\end{equation}
which minimizes the noiseless finite block least-squares criterion for
the realized samples and fading coefficients. QVPS and AQVPS use disturbance-detection AUC, while neural receivers use
evaluation MSE under matched training data, optimization settings, and
approximately matched real scalar parameter counts. Confidence intervals
are obtained by trial-level bootstrap resampling while preserving the
paired trial structure.

For the neural receiver experiments, evaluation MSE is reported per real
component as
\[
    \operatorname{MSE}
    =
    \frac{1}{4N_{\mathrm{test}}}
    \sum_{t=1}^{N_{\mathrm{test}}}
    \left\|
        \widehat{\bm x}_t-\bm x_t
    \right\|_2^2,
\]
where each \(\bm x_t\in\mathbb C^2\) contains four real components.

\begin{figure}[b!]
\centering
\includegraphics[width=\columnwidth]{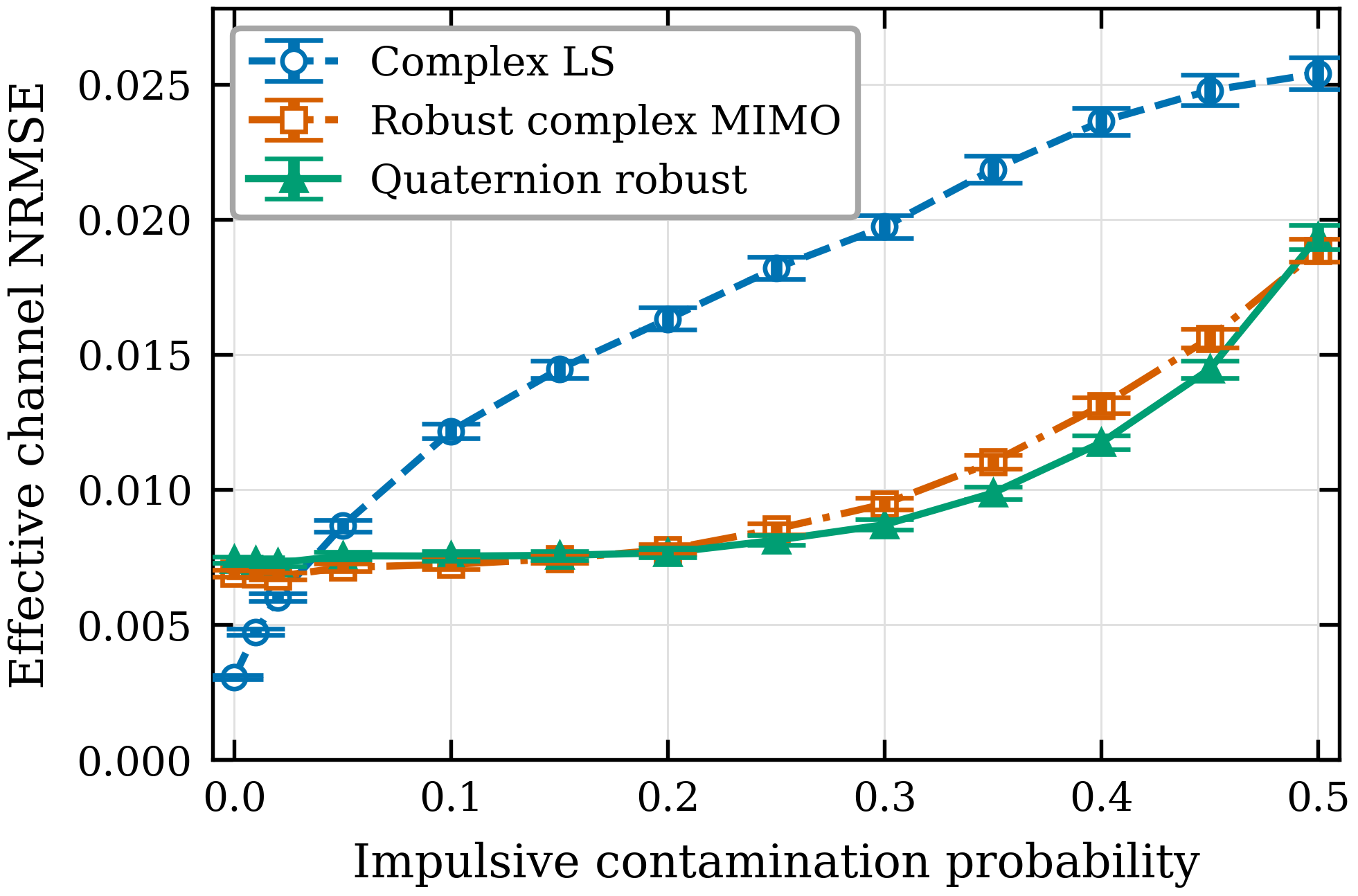}
\caption{Effective channel NRMSE versus impulsive contamination
probability. Complex LS has the lowest error in the clean regime,
whereas the robust estimators degrade more slowly with increasing
contamination.}
\label{fig:channel_nrmse}
\end{figure}

\section{Numerical Experiments}
\label{sec:results}

The experiments first examine robust effective-channel estimation as
the contamination probability increases. We then compare
QVPS and AQVPS under power-matched localized disturbances involving
I/Q impropriety, cross-polarization pseudo-covariance, and
cross-polarization covariance. Finally, matched real-, complex-, and
quaternion-valued neural receivers are evaluated under controlled
receiver impairments, with a widely linear complex receiver included
when conjugate dependence is present.

Unless stated otherwise, complex Gaussian samples with variance \(v\)
are generated as \(z=\sqrt{v/2}(a+j b)\), where
\(a,b\sim\mathcal N(0,1)\) are independent, so that
\(\mathbb E|z|^2=v\) and each real quadrature has variance \(v/2\).
All reported variances are normalized complex baseband quantities, not
shot-noise units.

The reference Gaussian regime evaluates receiver performance without
non-Gaussian contamination or structured receiver disturbances. The
real, complex, and quaternion neural receivers use matched training and
test data, common optimization settings, and approximately matched
real-scalar parameter counts. This provides the clean-data reference for
the robustness experiments that follow.

The complex neural receiver uses \texttt{torchcvnn}~\cite{Fix2025},
while the quaternion receiver follows the Hamilton-product construction
in~\cite{Parcollet2019} with the right-Hamilton layer of
Section~\ref{sec:qnr}. Neural comparisons use matched data,
optimization settings, and approximately matched real-scalar parameter
counts.

We use the same $4096$ revealed reference samples for Complex LMMSE and the
supervised receivers. We estimate the effective forward channel by LS and a
scalar proper-complex receiver-noise variance from the reference residuals.
We use a unit proper-complex modulation prior variance. Complex LMMSE uses
neither oracle channel information nor oracle noise information, and it uses
no separately optimized pilot sequence. Using the column-stacked reference matrices
\[
    \bm X_{\mathrm{ref}}
    =
    [\bm x_1,\ldots,\bm x_{N_{\mathrm{ref}}}],
    \qquad
    \bm Y_{\mathrm{ref}}
    =
    [\bm y_1,\ldots,\bm y_{N_{\mathrm{ref}}}],
\]
the Complex LMMSE baseline estimates the effective forward channel as
\begin{equation}
    \widehat{\bm A}
    =
    \bm Y_{\mathrm{ref}}
    \bm X_{\mathrm{ref}}^{H}
    \left(
        \bm X_{\mathrm{ref}}
        \bm X_{\mathrm{ref}}^{H}
    \right)^{-1}.
\end{equation}
The scalar complex noise variance is estimated as
\begin{equation}
    \widehat{\sigma}_{n}^{2}
    =
    \frac{
        \left\|
            \bm Y_{\mathrm{ref}}
            -
            \widehat{\bm A}\bm X_{\mathrm{ref}}
        \right\|_{F}^{2}
    }{
        2N_{\mathrm{ref}}
    }.
\end{equation}
With \(\bm C_x=\bm I_2\), the resulting receiver is
\begin{equation}
    \widehat{\bm x}_t
    =
    \widehat{\bm A}^{H}
    \left(
        \widehat{\bm A}\widehat{\bm A}^{H}
        +
        \widehat{\sigma}_{n}^{2}\bm I_2
    \right)^{-1}
    \bm y_t .
\end{equation}
The \(4096\)-sample reference block contains no impulsive contamination,
which is introduced only during evaluation.

\begin{figure*}[t!]
\centering
\subfloat[I/Q impropriety\label{fig:structured_iq}]
{\includegraphics[width=0.32\textwidth]{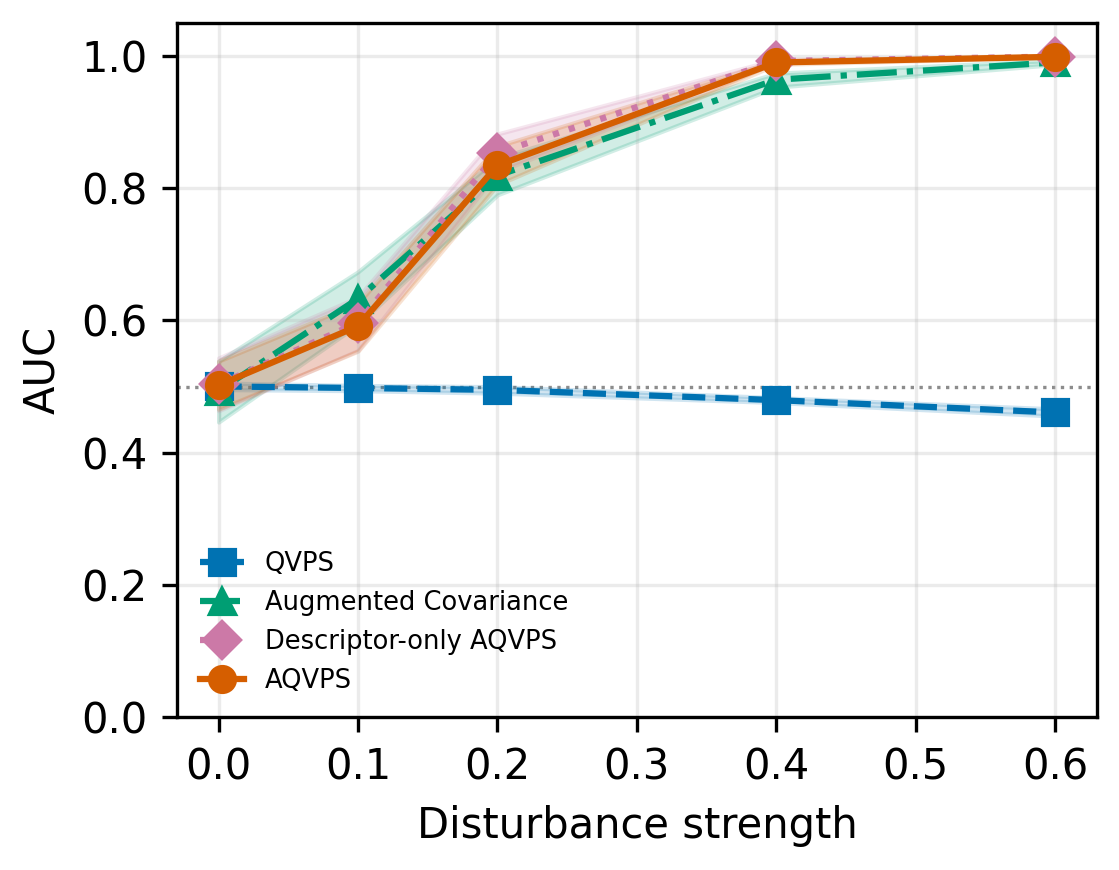}}
\hfil
\subfloat[Cross-polarization pseudo-covariance\label{fig:structured_pcov}]
{\includegraphics[width=0.32\textwidth]{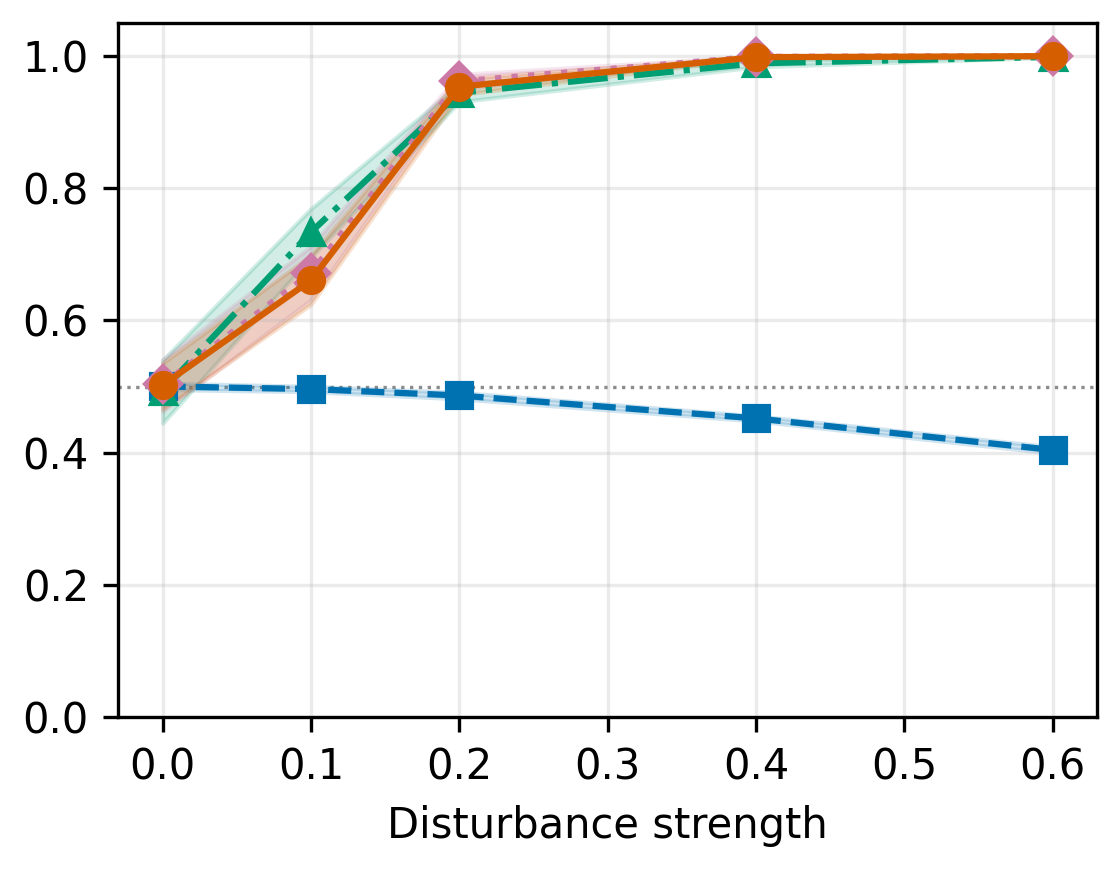}}
\hfil
\subfloat[Cross-polarization covariance\label{fig:structured_cov}]
{\includegraphics[width=0.32\textwidth]{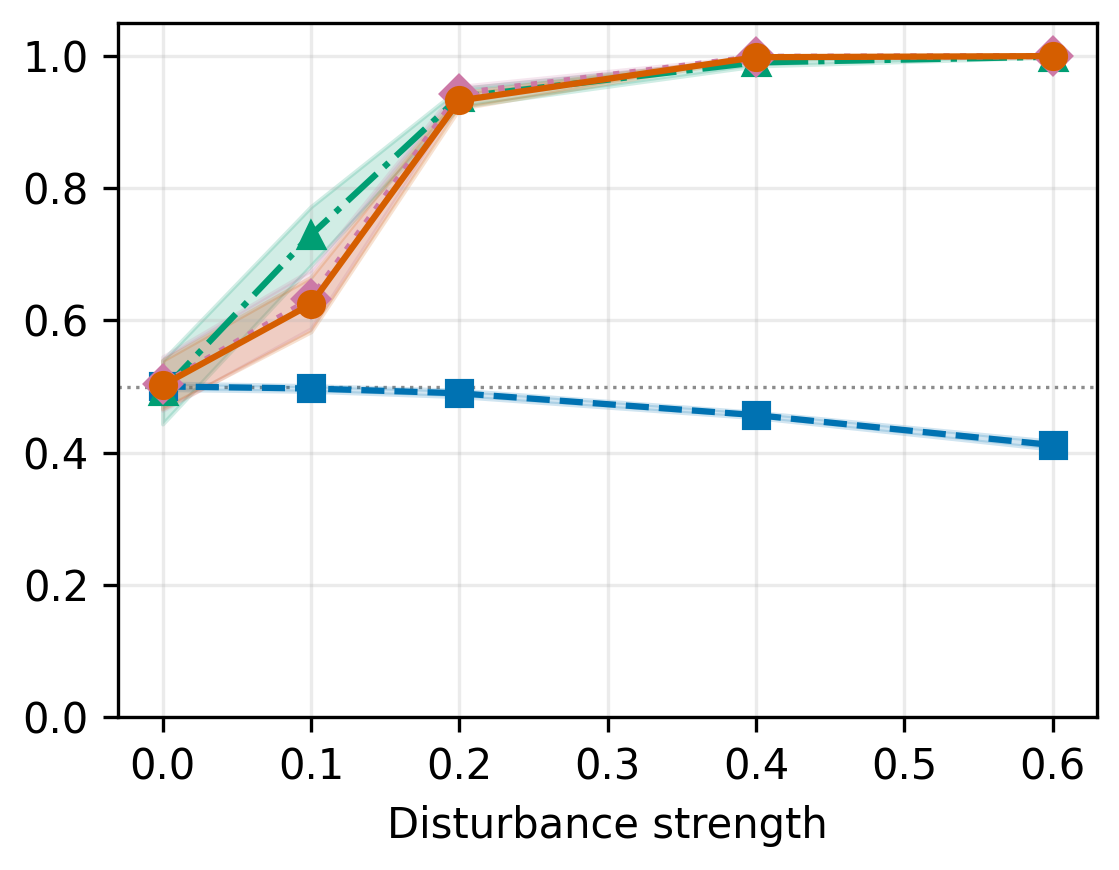}}
\caption{Disturbance-detection AUC for QVPS, AQVPS,
descriptor-only AQVPS, and a conventional augmented covariance
detector under (a) I/Q impropriety, (b) cross-polarization
pseudo-covariance, and (c) cross-polarization covariance. The augmented
covariance detector gives the highest mean AUC at $\alpha=0.1$, while
the descriptor-only AQVPS gives the highest mean AUC for the tested
$\alpha\geq0.2$ conditions.}
\label{fig:structured_auc}
\end{figure*}

\subsection{Non-Gaussian Contamination}

Robust receiver parameter estimation is evaluated as the contamination
probability increases. The comparison includes complex least squares,
robust complex estimation, and quaternion robust estimation. Performance
is measured using effective channel NRMSE. The paired Monte Carlo
experiments quantify the contamination level at which robust estimation
begins to offset the clean-data efficiency advantage of least squares.
We use a Bernoulli--Gaussian impulsive contamination model defined as:
\begin{equation}
    \bm n_t
    =
    \bm n_t^{(G)}
    +
    b_t\bm e_t,
\end{equation}
where
\begin{equation}
\begin{aligned}
    \bm n_t^{(G)}
    &\sim
    \mathcal{N}_{\mathrm C}(\bm 0,0.02\,\bm I),\\
    b_t
    &\sim
    \operatorname{Bernoulli}(\epsilon),\\
    \bm e_t
    &\sim
    \mathcal{N}_{\mathrm C}\!\left(
        \bm 0,
        12^2\times0.02\,\bm I
    \right).
\end{aligned}
\end{equation}

Fig.~\ref{fig:channel_nrmse} summarizes the resulting
effective channel estimation performance under impulsive contamination.
Complex LS gives the lowest NRMSE for $\epsilon\leq0.02$, while the
robust complex and quaternion estimators degrade more slowly as the
contamination probability increases. Robust complex estimation gives
the lower NRMSE for $\epsilon=0.05$--$0.15$. Quaternion robust
estimation gives the lower NRMSE for $\epsilon=0.20$--$0.45$, while
robust complex estimation again gives the lower NRMSE at $\epsilon=0.50$.
Paired bootstrap confidence intervals for the difference between the two
robust estimators exclude zero over these ranges.

\subsection{Structured Second-Order Disturbances}

QVPS and AQVPS are evaluated under localized disturbances involving
I/Q impropriety, cross-polarization pseudo-covariance, and
cross-polarization covariance. Each trial uses $N=3000$ receiver
samples, $L=64$ QVPS projection directions, and an AQVPS window length
of $W=151$. Results are obtained from $30$ paired Monte Carlo trials.

Before applying the structured disturbances, each trial is generated
from the uncontaminated fading model with
\[
    \bm x_t\sim\mathcal{N}_{\mathrm C}(\bm 0,4\bm I_2),
    \qquad
    \bm n_t\sim\mathcal{N}_{\mathrm C}(\bm 0,0.02\bm I_2).
\]
The pre-disturbance residual is formed relative to the block-average
effective channel as
\[
    \bm r_t
    =
    \left(
        \sqrt{T_t}
        -
        \frac{1}{N}\sum_{s=1}^{N}\sqrt{T_s}
    \right)
    \bm H\bm x_t
    +
    \bm n_t .
\]
Thus, the clean residual contains Gaussian receiver noise and residual
samplewise fading relative to the block-average effective channel.

The disturbances are applied to one central contiguous segment
\(\mathcal I\) containing \(25\%\) of the samples. Before power
normalization, the three disturbance families are
\begin{align}
\text{I/Q:}\quad
(\tilde r_X,\tilde r_Y)
&=
(r_X+\alpha\overline{r_X},r_Y),
\nonumber\\
\text{Pseudo-covariance:}\quad
(\tilde r_X,\tilde r_Y)
&=
(r_X+\alpha\overline{r_Y},r_Y+\alpha\overline{r_X}),
\nonumber\\
\text{Covariance:}\quad
(\tilde r_X,\tilde r_Y)
&=
(r_X+\alpha r_Y,r_Y+\alpha r_X).
\label{eq:structured_disturbances}
\end{align}
Each polarization is then power matched over \(\mathcal I\) according to
\begin{equation}
r'_p
=
\tilde r_p
\sqrt{
\frac{
\langle |r_p|^2\rangle_{\mathcal I}
}{
\max\!\left(
\langle|\tilde r_p|^2\rangle_{\mathcal I},
10^{-12}
\right)
}
},
\qquad
p\in\{X,Y\},
\end{equation}
so that discrimination is not driven by increased segment power. Performance is measured using disturbance detection AUC. These experiments test whether the local second-order descriptors used by AQVPS reveal receiver
anomalies that remain weakly expressed in the samplewise QVPS outlyingness.

As a conventional second-order baseline, we also use an augmented covariance
detector. For the augmented-covariance baseline, define
\[
    \bm v_t
    =
    \begin{bmatrix}
        r_{X,t} &
        r_{Y,t} &
        \overline{r_{X,t}} &
        \overline{r_{Y,t}}
    \end{bmatrix}^{T}.
\]
For the local window \(\mathcal W_t\),
\[
    \bm R_{a,t}
    =
    \frac{1}{|\mathcal W_t|}
    \sum_{s\in\mathcal W_t}
    \bm v_s\bm v_s^{H}.
\]
The corresponding detector score is
\[
    s_t
    =
    \frac{
        \left\|
            \bm R_{a,t}-\bm R_{a,0}
        \right\|_F
    }{
        \max\!\left(
            \left\|\bm R_{a,0}\right\|_F,
            \epsilon_{\mathrm{AQVPS}}
        \right)
    },
\]
where \(\bm R_{a,0}\) is the augmented residual covariance estimated
from the three clean calibration blocks.

Fig.~\ref{fig:structured_auc} shows the resulting disturbance
discrimination. Samplewise QVPS remains near or below chance for the
tested power-matched second-order disturbances. The conventional
augmented-covariance detector gives the highest mean AUC at the weakest
nonzero disturbance strength, $\alpha=0.1$, for all three disturbance
families. At $\alpha\geq0.2$, the descriptor-only AQVPS ablation gives
the highest mean AUC in the tested cases, while full AQVPS remains close
and approaches perfect discrimination at the larger disturbance
strengths. These results show that the gain is driven primarily by the
local second-order descriptors rather than the samplewise QVPS
outlyingness.

\begin{figure}[t!]
\centering
\includegraphics[width=\columnwidth]{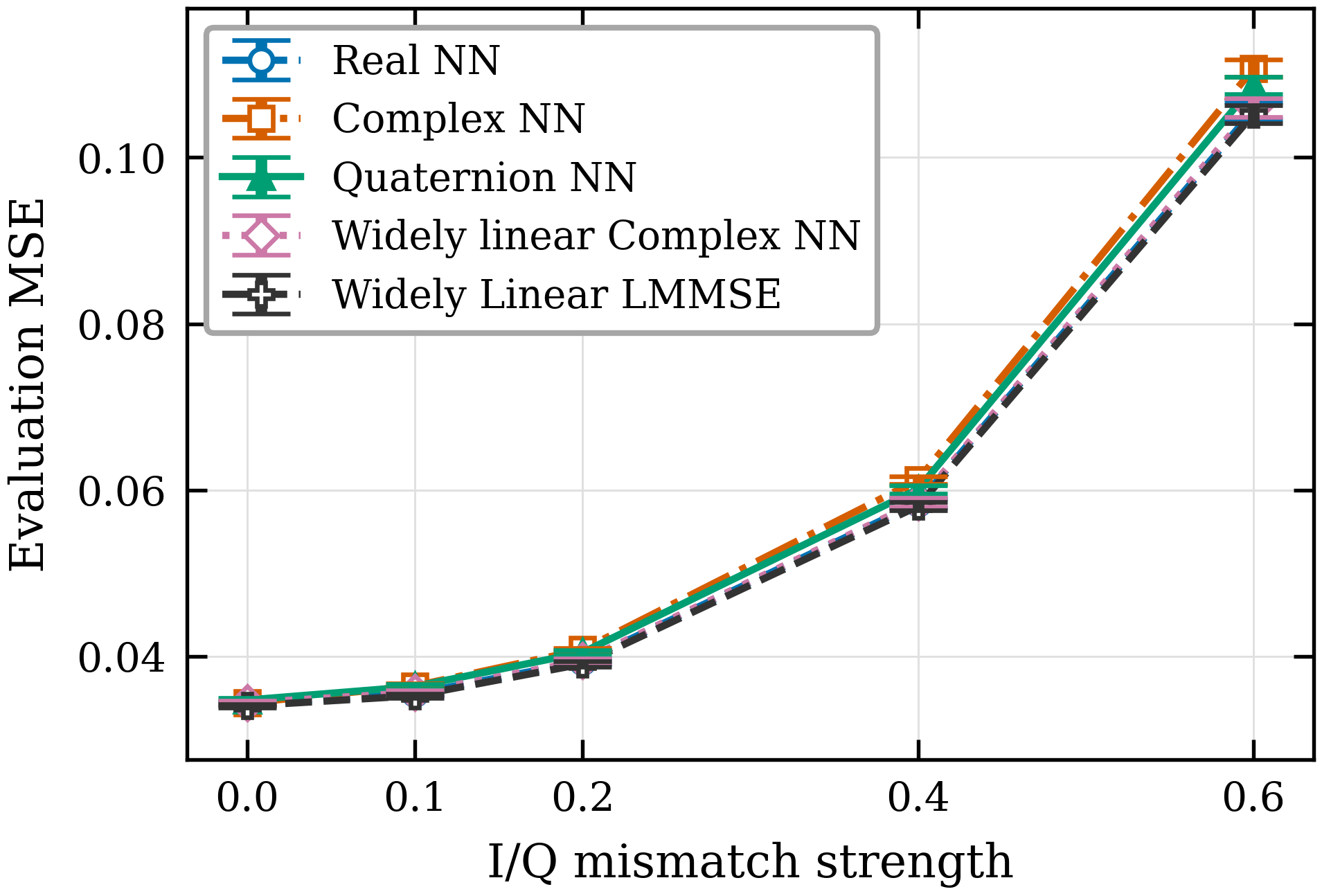}
\caption{Evaluation MSE versus I/Q mismatch strength for Widely Linear
LMMSE and matched Real NN, Complex NN, Quaternion NN, and widely linear
Complex NN receivers. Widely Linear LMMSE gives the lowest mean
evaluation MSE throughout the tested range. The Real NN and widely
linear Complex NN remain close to this baseline, while the
right-Hamilton Quaternion NN degrades more slowly than the strictly
Complex NN as the mismatch increases.}
\label{fig:iq_mismatch}
\end{figure}

\subsection{Neural Receiver Impairments}

The neural receiver experiments use $20$ paired Monte Carlo trials with
common training and test data across the compared receivers. For
the impulsive-contamination and \textrm{SNR} experiments, the networks are trained
on Gaussian-clean data and impulsive contamination is introduced only in
the test data. For the I/Q-mismatch experiment, the networks are
trained and evaluated at the same specified mismatch level. The Real NN,
Complex NN, and Quaternion NN use common optimization settings and
approximately matched real-scalar parameter counts.
\begin{figure}[b!]
\centering
\includegraphics[width=\columnwidth]{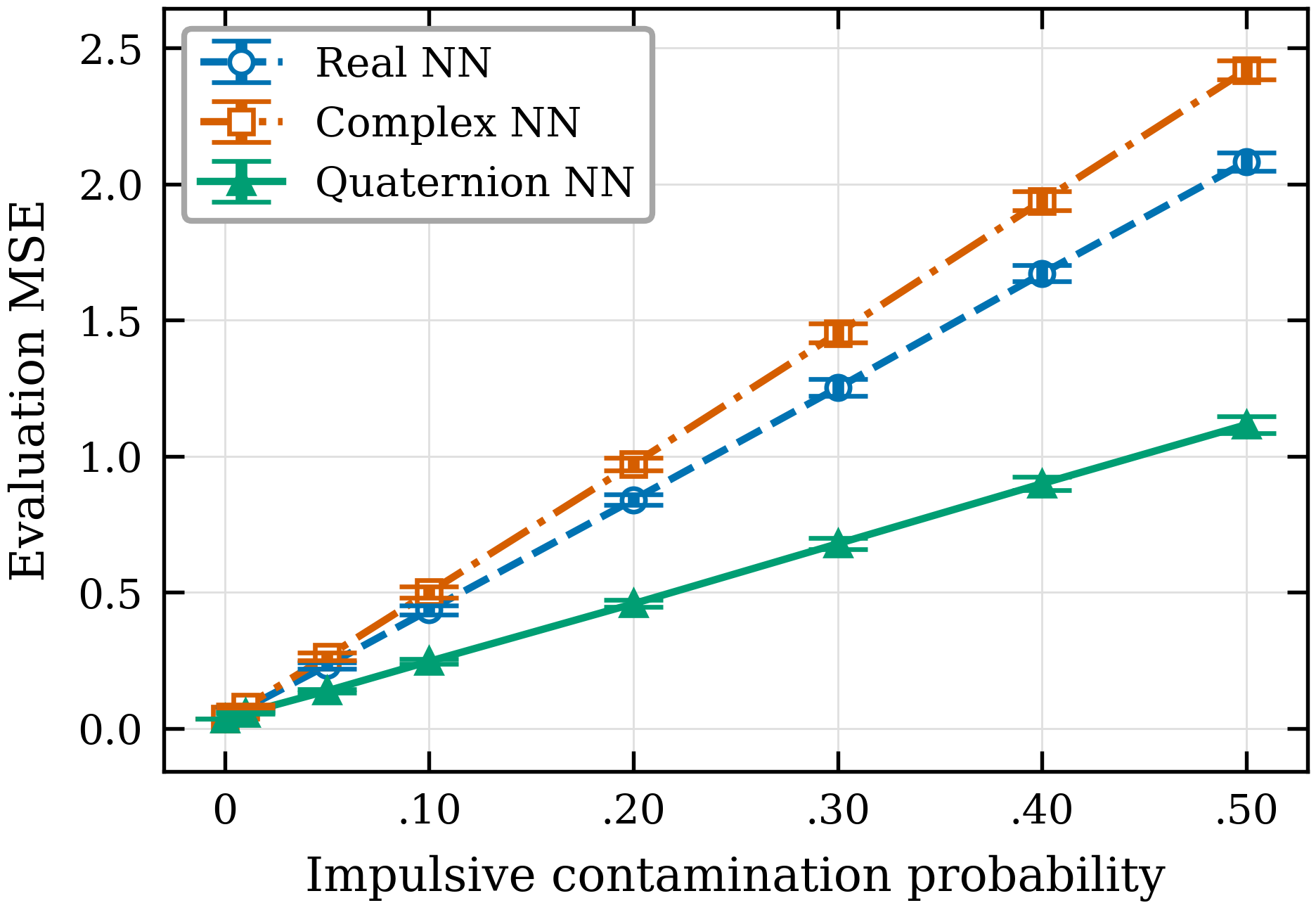}
\caption{Evaluation MSE versus impulsive contamination probability for
Real NN, Complex NN, and Quaternion NN. The quaternion receiver incurs
a small clean-data efficiency penalty but degrades substantially more
slowly as the contamination probability increases.}
\label{fig:impulse_mse}
\end{figure}

\begin{figure*}[t!]
    \centering
    \subfloat[$\epsilon=0$\label{fig:snr_eps000}]
    {\includegraphics[width=0.45\textwidth]{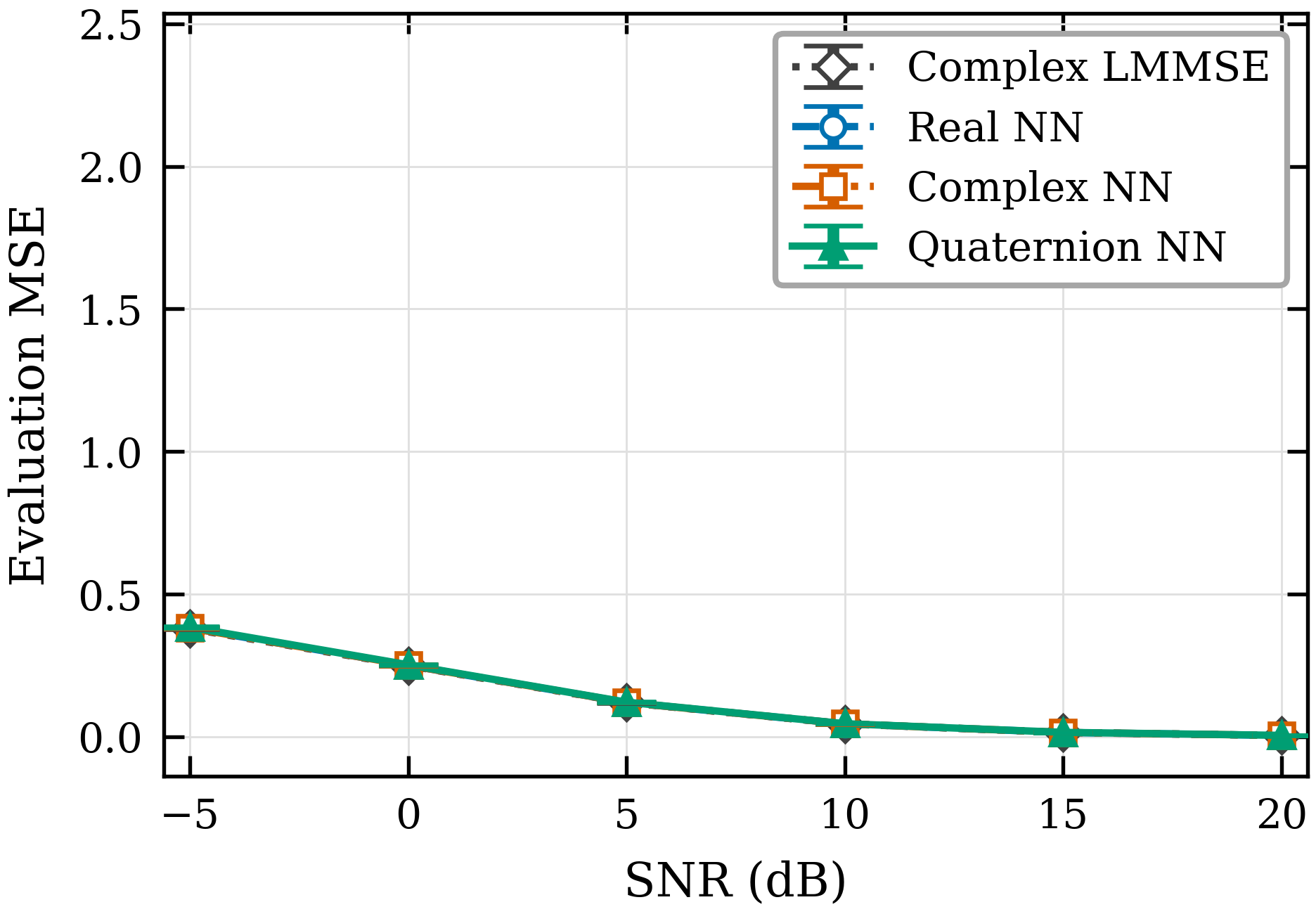}}
    \hfil
    \subfloat[$\epsilon=0.15$\label{fig:snr_eps015}]
    {\includegraphics[width=0.45\textwidth]{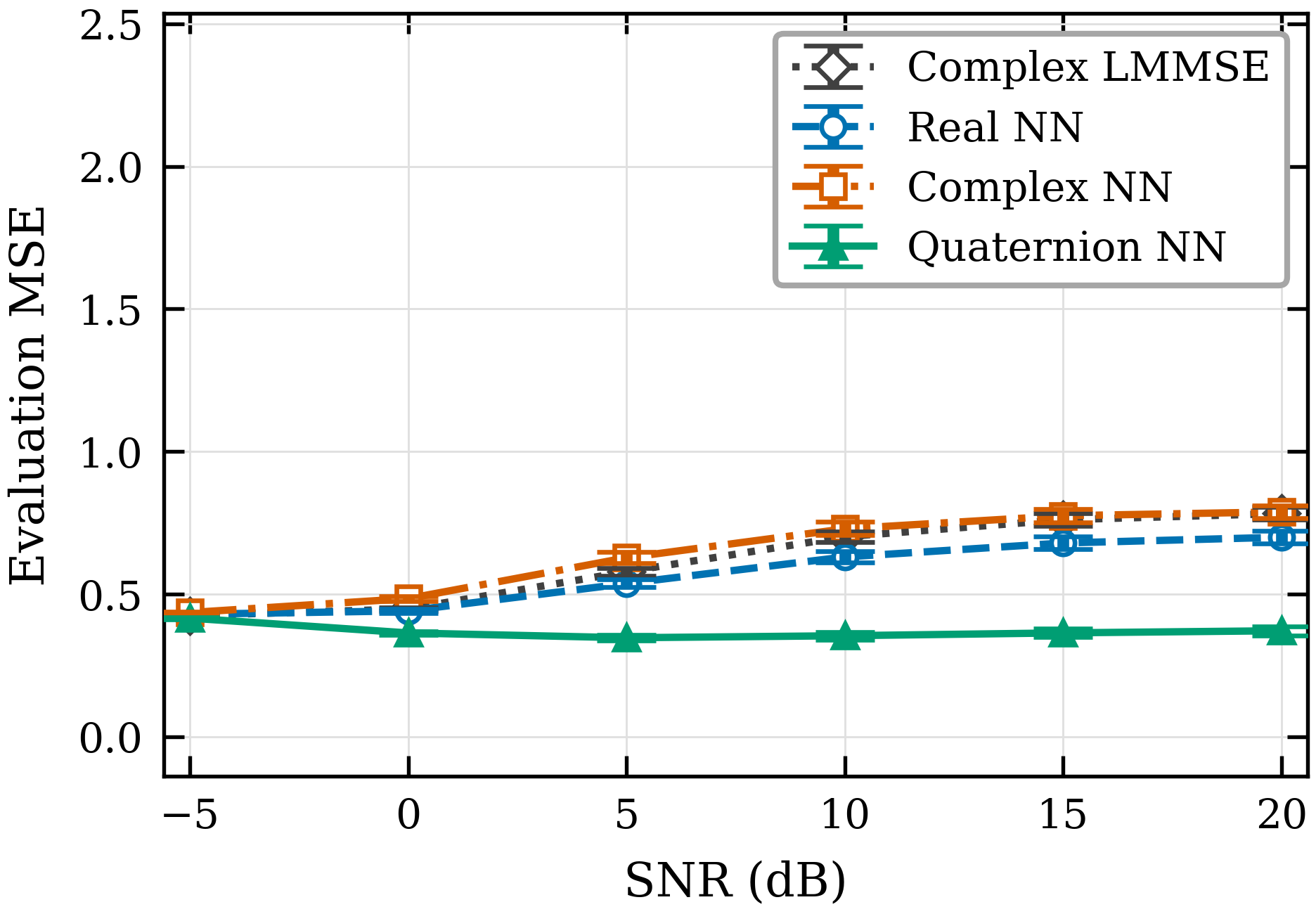}}
    \\
    \subfloat[$\epsilon=0.30$\label{fig:snr_eps030}]
    {\includegraphics[width=0.45\textwidth]{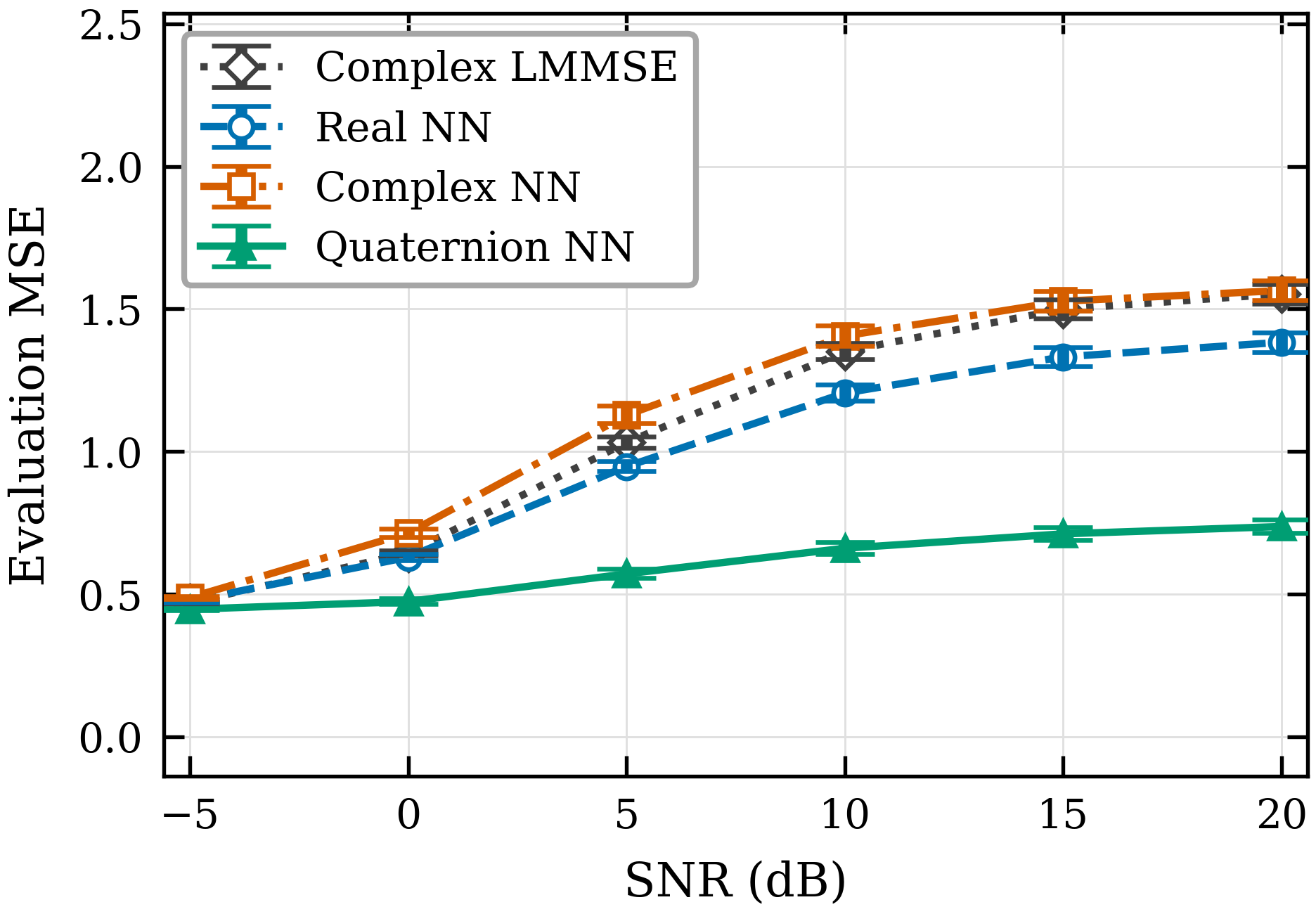}}
    \hfil
    \subfloat[$\epsilon=0.45$\label{fig:snr_eps045}]
    {\includegraphics[width=0.45\textwidth]{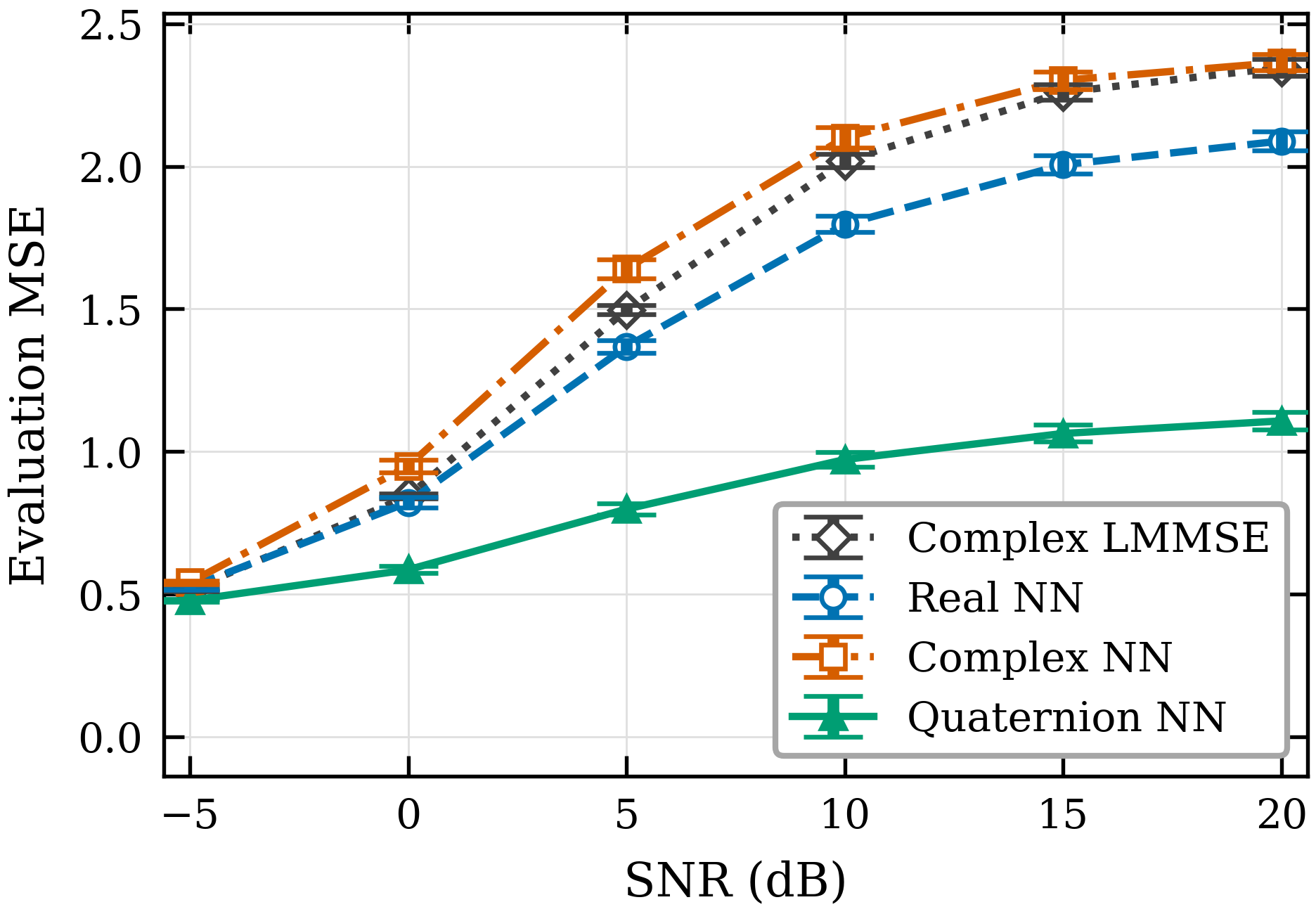}}
    \caption{Evaluation MSE versus \textrm{SNR} for Complex LMMSE and matched
    Real NN, Complex NN, and Quaternion NN receivers under clean conditions
($\epsilon=0$) and under impulsive contamination
($\epsilon=0.15,0.30,0.45$). Complex LMMSE has the lowest
evaluation MSE throughout the clean Gaussian \textrm{SNR} sweep. Among the
compared receivers with their nominal activations, the Quaternion NN has the
lowest mean evaluation MSE throughout the tested contaminated \textrm{SNR}
sweeps. As the nominal Gaussian-noise variance decreases, the fixed
impulsive disturbance becomes dominant, producing the LMMSE high-\textrm{SNR}
contamination floor and increase.}
    \vspace{1mm}
    \label{fig:snr_mse}
\end{figure*}

Define the post-channel, pre-impairment signal as
\[
    \bm z
    =
    \sqrt{\eta}\,
    \bm H(\theta,\phi,\kappa)\bm x .
\]
For the I/Q mismatch experiment, both polarizations are transformed as
\begin{equation}
    \bm y
    =
    \frac{\bm z+\alpha\overline{\bm z}}
    {\sqrt{1+\alpha^{2}}}
    +\bm n,
\end{equation}
where $\alpha\in\{0,0.1,0.2,0.4,0.6\}$ is fixed within each condition and $\bm n$ denotes the additive receiver noise.

For this conjugate-dependent model, Widely Linear LMMSE uses the same
$4096$ revealed reference samples as the supervised receivers. Defining
\begin{equation}
    \bm y_{\mathrm{aug},t}
    =
    \begin{bmatrix}
        \bm y_t\\
        \overline{\bm y_t}
    \end{bmatrix},
    \qquad
    \widehat{\bm x}_t
    =
    \widehat{\bm C}_{x y_{\mathrm{aug}}}
    \widehat{\bm C}_{y_{\mathrm{aug}} y_{\mathrm{aug}}}^{-1}
    \bm y_{\mathrm{aug},t},
\end{equation}
the two covariance matrices are estimated empirically from that reference
block only. This provides a non-neural, reference-estimated approximation to the
Gaussian conditional-mean receiver for the widely linear I/Q-mismatch
model and uses neither channel nor covariance information from the
evaluation block.

Matched real-, complex-, and quaternion-valued neural receivers are
evaluated under controlled dual-polarization receiver impairments,
including I/Q mismatch and impulsive contamination. The right-Hamilton
quaternion layers defined in Section~\ref{sec:qnr} are used for the
Quaternion NN. Their multiplication convention is algebraically
compatible with the structured Jones family used for polarization
mixing. Under I/Q mismatch, they remain a structured receiver bias,
whereas the widely linear complex receiver represents conjugate
dependence explicitly.

Fig.~\ref{fig:iq_mismatch} shows the evaluation MSE under increasing
I/Q mismatch. Widely Linear LMMSE gives the lowest mean evaluation MSE
throughout the tested range. The Real NN and widely linear Complex NN
remain close to this baseline, whereas the Quaternion NN degrades more
slowly than the strictly Complex NN as the mismatch increases. 

Under impulsive contamination, the relative behavior changes more
substantially. As shown in Fig.~\ref{fig:impulse_mse}, the Quaternion NN
incurs a small clean-data efficiency penalty, but its evaluation error
grows considerably more slowly as the contamination probability
increases. At a contamination probability of $0.50$, the mean
evaluation MSE is approximately $1.116$ for the Quaternion NN, compared
with $2.082$ and $2.419$ for the Real NN and Complex NN, respectively. These experiments quantify the degradation of receivers calibrated on
uncontaminated observations.

For the \textrm{SNR} sweeps, the post-channel, pre-noise signal power is
\begin{equation}
    P_{\mathrm{sig}}
    =
    \eta\left(
        \cos^{2}\theta
        +
        \kappa^{2}\sin^{2}\theta
    \right),
\end{equation}
and the nominal Gaussian-noise variance is
\begin{equation}
    \sigma_n^{2}
    =
    \frac{P_{\mathrm{sig}}}
    {10^{\mathrm{SNR}_{\mathrm{dB}}/10}}.
\end{equation}
The impulsive disturbance variance is held fixed at
$\sigma_{\rm impulse}^2=2.88$ while the nominal Gaussian-noise variance
changes with \textrm{SNR}.
Fig.~\ref{fig:snr_mse} compares the receivers across \textrm{SNR}.
Complex LMMSE has the lowest evaluation MSE throughout the clean Gaussian
\textrm{SNR} sweep. Among the compared receivers with their nominal activations, the
Quaternion NN has the lowest mean evaluation MSE throughout the
tested contaminated \textrm{SNR} sweeps. The fixed impulsive disturbance becomes
dominant as the nominal Gaussian-noise variance decreases, producing the
Complex LMMSE high-\textrm{SNR} contamination floor and increase.
\begin{figure}[t!]
\centering
\includegraphics[width=\columnwidth]{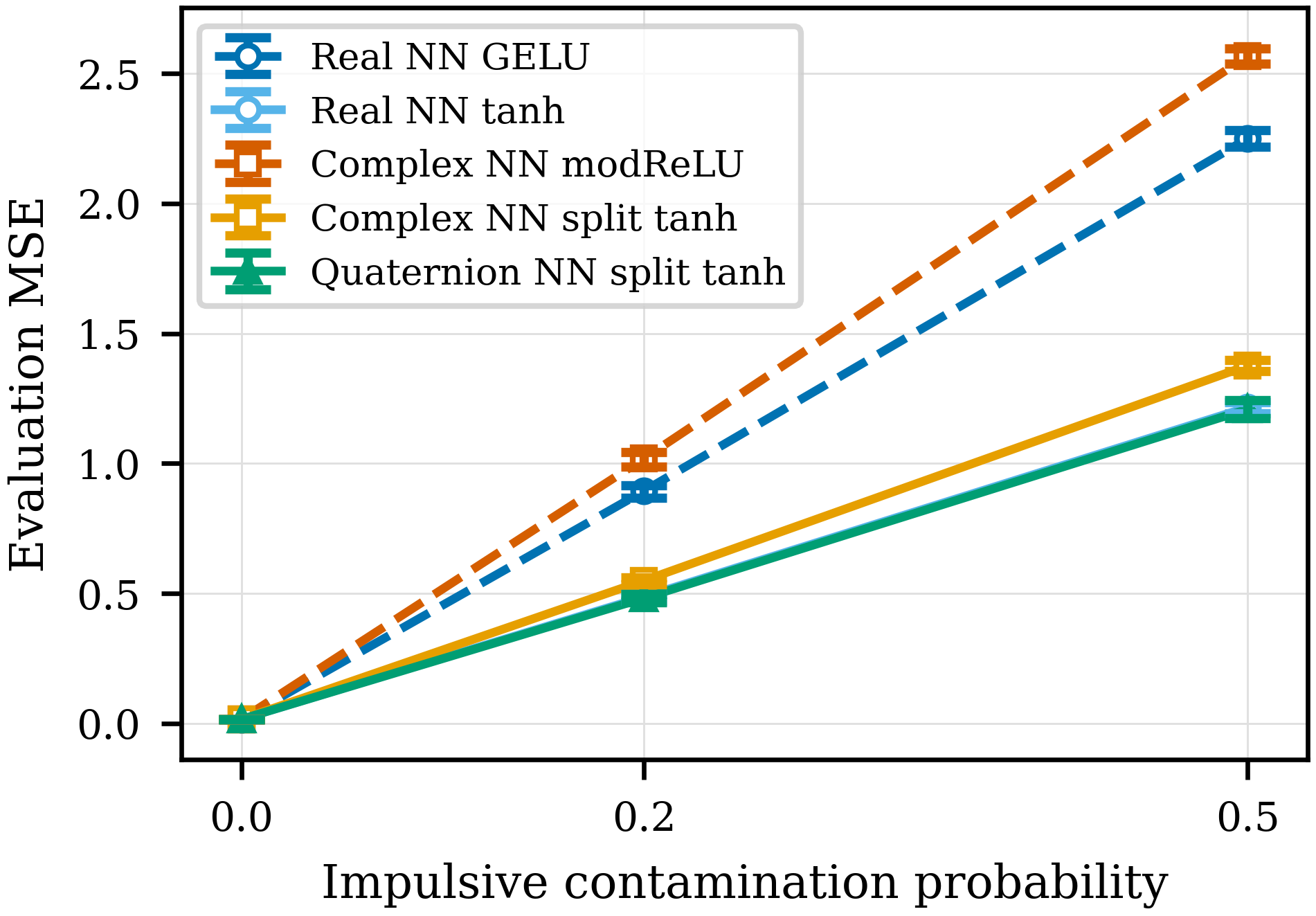}
\caption{Activation-matched ablation under impulsive contamination at
$15$~dB \textrm{SNR}. Evaluation MSE is shown for the nominal Real NN and Complex NN
architectures, their bounded activation variants, and the right-Hamilton
Quaternion NN. Error bars indicate paired-bootstrap $95\%$ confidence
intervals.}
\label{fig:ablation}
\end{figure}

To separate the effect of quaternion structure from that of the bounded
activation, we performed an activation-matched ablation at $15$~dB \textrm{SNR},
shown in Fig.~\ref{fig:ablation}. The ablation uses $10$ paired Monte Carlo
trials with $\epsilon\in\{0,0.20,0.50\}$. In addition to the nominal Real NN and Complex NN, the Real NN is
evaluated with $\tanh$ activation and the Complex NN with split-$\tanh$
activation, while the right-Hamilton Quaternion NN retains its
split $\tanh$ activation. All other training and test
conditions are unchanged. The bounded activation
variants substantially reduce the degradation caused by impulsive
contamination, indicating that activation saturation accounts for a large
part of the robustness observed for the nominal Quaternion NN. Under activation matching, the paired-bootstrap intervals for the
Quaternion NN minus Real NN with $\tanh$ include zero at
$\epsilon=0.20$ and $0.50$, while the Quaternion NN gives lower MSE
than the Complex NN with split-$\tanh$ under contamination.

\section{Complexity and Implementation}
\label{sec:complexity}

Let $N$ denote the number of receiver samples, $L$ the number of
quaternion projection directions, and $P$ the number of trainable
real-scalar neural parameters. QVPS has dominant scoring cost
$O(NL)$ per robust reweighting pass, in addition to robust location
and scale estimation for each projection direction. AQVPS adds the
computation of local second-order descriptors and their robust location
and scatter estimates.

Neural receiver training is dominated by forward and backward passes
and therefore scales with the number of training samples and trainable
parameters. The real, complex, and quaternion receivers use
approximately matched real-scalar parameter counts. Quaternion layers
nevertheless impose Hamilton-structured parameter sharing, so equal
parameter counts do not imply equal functional flexibility. The
structured Jones receiver uses the right-Hamilton multiplication
convention of Section~\ref{sec:qnr}.

\section{Security Relevance and Limitations}
\label{sec:security}

The proposed receiver is security-relevant because receiver estimates
affect residual noise, covariance structure, and other quantities used in
parameter estimation. Receiver robustness, however, is not equivalent to a
CV-QKD security proof. A learned or robust receiver may reduce
estimation bias in simulation while still requiring separate
protocol-specific security analysis before deployment. Accordingly,
the quantities reported in this paper are interpreted as
receiver-level performance metrics and security-relevant proxies rather
than as certified secret-key rates. Robust downweighting does not classify
a disturbance as trusted or untrusted and cannot be substituted into a
security analysis.

The quaternion formulation is not assumed to be uniformly superior to
real- or complex-valued processing. When robust complex MIMO or simpler
real-valued methods achieve comparable performance, those regimes are
reported explicitly. Such results help identify when quaternion
structure provides a measurable advantage and when simpler receiver
models are sufficient.

\section{Conclusion}
\label{sec:conclusion}

This paper developed a robust quaternion receiver methodology for
dual-polarization free-space quantum communications. By encoding the
two complex heterodyne polarization streams as quaternion-valued
samples, the receiver preserves their joint polarization and quadrature
structure. The proposed methodology combines quaternion neural processing with
QVPS and AQVPS diagnostics for samplewise and structure-aware
disturbance assessment.

The numerical results show that robust complex and quaternion estimation
reduce the effect of non-Gaussian contamination relative to conventional
complex least squares. Local second-order descriptors identify the tested
localized disturbances involving I/Q impropriety, cross-polarization
pseudo-covariance, and cross-polarization covariance that are weakly
expressed in samplewise QVPS scores. A descriptor-only ablation slightly
outperforms full AQVPS over the tested nonzero disturbance strengths,
showing that the structured-disturbance gain is driven primarily by the
local second-order descriptors rather than the samplewise QVPS coordinate.
Complex LMMSE has the lowest evaluation MSE throughout the clean Gaussian \textrm{SNR} sweep, whereas,
among the compared receivers with their nominal activations, the Quaternion
NN has the lowest mean evaluation MSE throughout the tested contaminated
\textrm{SNR} sweeps. For the neural receiver, the right-Hamilton construction
preserves the multiplication convention of the structured Jones mapping used
in the polarization mixing experiment.

For the nominal neural receiver activations, the Quaternion NN degrades
more slowly than the Real NN and Complex NN under impulsive
contamination. However, the activation-matched ablation shows that
bounded activation accounts for a large part of this robustness
difference. The paired difference intervals between the Quaternion NN
and the Real NN with tanh include zero at the tested contaminated
conditions, while the Quaternion NN gives lower MSE than the Complex NN
with split-tanh under contamination. Thus, quaternion processing is not
uniformly superior to real- or complex-valued alternatives. Its value
depends on the receiver impairment, estimator, and model structure. The
proposed methods should therefore be interpreted as receiver-level
robustness and estimation tools. Their use in a deployed CV-QKD system
would require separate protocol-specific security analysis.

\IEEEtriggeratref{12}
\bibliographystyle{IEEEtran}
\bibliography{IEEEabrv,quartz}

\end{document}